\def\grd{^{\protect\raisebox{-5pt}{$\circ$}}_{\protect\raisebox{3pt}{\,.}}}
\def\cross{\bds\times}

\def\cross{\bds\times}

\def\nk{n_{\rm b}}

\def\rfr#1{Equation\,(\ref{#1})}
\def\rfrs#1#2{Equations\,(\ref{#1})-(\ref{#2})}
\def\Rfr#1{Equation\,(\ref{#1})}
\def\Rfrs#1#2{Equations\,(\ref{#1})-(\ref{#2})}

\def\dert#1#2{\frac{{{\textrm{d}}}{#1}}{{{\textrm{d}}}{#2}}}
\def\virg#1{``#1"}

\def\eqi{\begin{equation}}
\def\eqf{\end{equation}}
\def\eqia{\begin{eqnarray}}
\def\eqfa{\end{eqnarray}}

\def\rp#1#2{{#1\over#2}}
\def\lb#1{\label{#1}}

\def\bds#1{\boldsymbol{#1}}
\def\ton#1{\left(#1\right)}
\def\qua#1{\left[#1\right]}
\def\grf#1{\left\{#1\right\}}

\documentclass[onecolumn]{aastex}

\usepackage{morefloats}
\usepackage[title]{appendix}
\usepackage{hyperref,textcomp}
\usepackage{booktabs}
\usepackage[table,xcdraw]{xcolor}
\usepackage{multirow}
\usepackage{rotating,tabularx}
\usepackage{float}
\usepackage{enumerate}
\usepackage{rotating}
\usepackage[polutonikogreek,english]{babel}
\usepackage{amsmath,starfont,textgreek,w-greek,wasysym}
\usepackage[flushleft]{threeparttable}
\usepackage{amsthm}
\usepackage{amscd,lineno}
\usepackage{amssymb,dsfont}
\usepackage{graphicx,epsfig}
\usepackage{txfonts}
\usepackage{xr-hyper}
\RequirePackage{color}

\newcommand{\emaila}{lorenzo.iorio@libero.it}

\allowdisplaybreaks[1]

\begin{document}

\title{\textcolor{black}{The effect of post-Newtonian spin precessions on the evolution of exomoons' obliquity}}

\shortauthors{L. Iorio}

\author{Lorenzo Iorio\altaffilmark{1} }
\affil{Ministero dell'Istruzione, dell'Universit\`{a} e della Ricerca
(M.I.U.R.)
\\ Viale Unit\`{a} di Italia 68, I-70125, Bari (BA),
Italy}

\email{\emaila}

\begin{abstract}
Putative natural massive satellites (exomoons) has gained increasing attention, where they orbit Jupiter-like planets within the habitable zone of their host main sequence star. An exomoon is expected to move within the equatorial plane of its host planet, with its spin ${\bds S}_\mathrm{s}$ aligned with its orbital angular momentum $\boldsymbol L$ which, in turn, is parallel to the planetary spin ${\boldsymbol S}_\mathrm{p}$. If, in particular, the common tilt of such angular momenta to the satellite-planet ecliptic plane, assumed fixed, has certain values, the  latitudinal irradiation  experienced on the exomoon from the star may allow it to sustain life as we know it, at least for certain orbital configurations.  An Earth--analog (similar in mass, \textcolor{black}{radius, oblateness} and obliquity) is considered, which orbits within $5-10$ planetary radii $R_\mathrm{p}$ from its Jupiter-like host planet. The de Sitter and Lense--Thirring spin precessions due to the general relativistic post-Newtonian (pN) field of the host planet have an impact on an exomoon's habitability for a variety of different initial spin-orbit configurations.  Here, I show it by identifying long--term  variations in the satellite's obliquity $\varepsilon_\mathrm{s}$, where variations can be $\lesssim 10^\circ-100^\circ$, depending on the initial spin-orbit configuration, with a timescale of $\simeq 0.1-1$ million years. Also the satellite's quadrupole mass moment $J_2^\mathrm{s}$ induces obliquity variations which are faster than the pN ones, but do not cancel them.
\end{abstract}

%{
%\textit{Unified Astronomy Thesaurus concepts}:\,Natural satellites (Extrasolar)\,(483); Exoplanets\,(498); Astrobiology\,(74); General relativity\,(641); %Relativistic mechanics\,(1391)
%}

\section{Introduction}
In investigating the possibility that alien worlds, extrasolar planets
%\footnote{See, e.g., http://phl.upr.edu/projects/habitable-exoplanets-catalog and http://exoplanet.eu/ on the Internet.}
\citep{2010exop.book.....S,2012NewAR..56....1L,2018haex.bookE....D,2018exha.book.....P} and related environments, may host and sustain known (and unknown) forms of life and, possibly, civilizations \citep{2017ARA&A..55..433K,2018NatAs...2..432S,2018AsBio..18..663S,2020Univ....6..130I}, it is of crucial importance to assess the physical conditions posing tight constraints. In this framework,  I will look at a novel scenario, \textcolor{black}{where} Einstein's General Theory of Relativity \citep[GTR;\,][]{2016Univ....2...23D,2017grav.book.....M}\textcolor{black}{, along with other classical effects,} may have a direct, macroscopic impact on life and its long-term sustainability.

Natural satellites of Jupiter--like gas giants, or exomoons \citep{2002ApJ...575.1087B,2006MNRAS.373.1227D,2014AsBio..14..798H,2015IJAsB..14..191S}, could be habitable if the host planet orbits within the habitable zone of its main sequence star \citep{1997Natur.385..234W,Kaltenegger_2010,Hel12,2013MNRAS.432.2994F,2013AsBio..13...18H,2013ApJ...774...27H,2014AsBio..14..798H,2017A&A...601A..91D,2017MNRAS.472....8Z,2018ApJ...860...67H,Forgan019,2019ApJ...887..261M,2020IJAsB..19..210L,2020A&A...636A..50T}.
\textcolor{black}{In the following, quantities pertaining the exomoon, its host planet and the star are labeled with s, p, and S, respectively.} Exomoons' mass $M_\mathrm{s}$ should be $0.25\,M_\oplus\lesssim M_\mathrm{s} \lesssim2\,M_\oplus$ to sustain life over a billion-year timescale \citep{2013AsBio..13...18H}. According to \citet{2010ApJ...714.1052S}, their actual formation around extrasolar giant planets is possible.
\textcolor{black}{In general, different mechanisms of formation of extrasolar planets's satellites have been proposed so far \citep{2016AstRv..12...24B}, including planet-planet collision, able to create satellites around rocky or icy planets \citep{2017MNRAS.466.4868B,2020MNRAS.492.5089M}, and co-accretion and  capture which should lead to gas giants' exomoons \citep{2014AsBio..14..798H,2016AstRv..12...24B}.}
Still unconfirmed exomoons candidates exist \citep{2020MNRAS.tmp.3526F,2018SciA....4.1784T}. The project Hunt for Exomoons with Kepler (HEK) was the most important effort aimed to detect exomoons to date \citep{2012ApJ...750..115K,2013ApJ...770..101K,2013ApJ...777..134K,2018AJ....155...36T}. Searches for exomoons started in 2009 \citep{2009MNRAS.392..181K,2009MNRAS.396.1797K,2009MNRAS.400..398K}, after theoretical investigations about such a possibility \citep{1999A&AS..134..553S,2007A&A...464.1133C}. \textcolor{black}{Techniques to be used in exomoons' detection are transit timing variations (TTVs), transit duration variations (TDVs), and apparent planetary transit radius variations (TRVs) \citep{2020A&A...638A..43R}; according to \citet{2020A&A...638A..43R}, TRVs could be a more promising means to identify exomoons in large exoplanet surveys.}

Such exomoons could be tidally locked to their parent planet but not to the host star, and moving  in the planetary equatorial plane due to  tidal evolution \citep{2011ApJ...736L..14P,2013AsBio..13...18H}. Moreover, the satellite's spin ${\bds S}_\mathrm{s}$ should be parallel to the
orbital angular momentum $\bds L$ of the planetocentric motion \citep{2013AsBio..13...18H}. Thus, the exomoon should have the same obliquity
with respect to the circumstellar orbit as the planetary spin ${\bds S}_\mathrm{p}$ \citep{2013AsBio..13...18H}, so that it could experience seasons if the equator of the host planet is tilted against the ecliptic plane \citep{2013AsBio..13...18H}. This scenario is plausible because previous studies have shown that exomoons can maintain significant obliquities on large timescales \citep{2011A&A...528A..27H,2013AsBio..13...18H}.

One of the key parameters for the long-term habitability of an astronomical major body  is the axial tilt $\varepsilon$, or obliquity, of its spin to its orbital plane, and its long-term stability over the \ae ons. In the case of a star-planet scenario, the planetary axial tilt $\varepsilon$ to the ecliptic plane is crucial for the latitude-dependent insolation received from the host star \citep{1993Natur.361..615L,1997Icar..129..254W,2004A&A...428..261L,2014AsBio..14..277A,2015P&SS..105...43L,
2017arXiv171008052Q,2017ApJ...844..147K,2018AJ....155..237S,2019arXiv191108431Q}. Indeed, variations in obliquity, meant as difference between its extreme values, drive changes in planetary climate. If the obliquity variations  are rapid and/or large, the resulting climate shifts can be commensurately severe \textcolor{black}{\citep{2004Icar..171..255A}}. As far as the Earth is concerned, its obliquity  changes slowly with time from $\simeq 22\grd1$ to $24\grd5$, undergoing an oscillation cycle with amplitude $\lesssim 2\grd4$ in about $41,000\,\mathrm{yr}$ \citep{2019arXiv191108431Q}. The value of the Earth's obliquity  impacts the seasonal cycles and its long-term variation affects the terrestrial climate \citep{Milan1941}, as deduced from geologic records \citep{Kerr1987,1995GeoJI.121...21M,1999E&PSL.174..155P}. For an exomoon, tidal heating, reflected light by the planet, and the planet's own infrared irradiation affects the total energy budget in addition to the direct stellar radiation \citep{2013AsBio..13...18H}. Thus, it is arguable that the long-term changes of the obliquity \textcolor{black}{$\vartheta_\mathrm{s}$} relative to the circumplanetary orbital plane will also affect the climate of the exomoon \textcolor{black}{in addition to those of the obliquity $\varepsilon_\mathrm{s}$ with respect to the ecliptic plane}.

The purpose of this paper is to show that GTR may \textcolor{black}{concur to} directly affect the habitability of an exomoon through the gravitoelectric de Sitter \citep{1916MNRAS..77..155D,1918KNAB...27.214S,1921KNAB...23..729F}  and gravitomagnetic Lense-Thirring\footnote{Such a denomination for the pN spin precession induced by the primary's angular momentum has become of common use, despite it was discovered by Pugh and Schiff in the sixties of the twentieth century.} \citep{Pugh59,Schiff60} rates of change of its spin ${\bds S}_\mathrm{s}$ relative to the ecliptic \textcolor{black}{along with other possible classical precessions due to, e.g., tidal friction, quadrupole mass moments $J^\mathrm{p}_2,\,J^\mathrm{s}_2$ of the planet and the satellite, 3rd-body effects due to the distant star and other major bodies in the system}. They are induced by the post-Newtonian (pN) static and stationary components of the gravitational field of its parent planet \citep{2013grsp.book.....O,2014grav.book.....P}. To the benefit of a reader not acquainted with GTR, the pN expansion is one of the most successful and famous approximation schemes that have been developed in the past years for solving the fully nonlinear Einstein's equations to describe motions of arbitrary shaped, massive bodies \citep{1997PThPS.128..123A,2003Blanchet,2018tegp.book.....W}. Furthermore,  the terms \virg{gravitoelectric} and \virg{gravitomagnetic} have nothing to do with electric charges and currents, referring, instead, to the formal resemblance of the linearized Einstein field equations of GTR, valid in the slow-motion and weak-field approximation, with the linear Maxwellian equations of electromagnetism \citep{Thorne86,2001rfg..conf..121M,2001rsgc.book.....R}. Both such effects were successfully measured some years ago in the dedicated spaceborne experiment Gravity Probe B (GP-B) with four artificial gyroscopes orbiting in the field of the Earth \citep{2011PhRvL.106v1101E,2015CQGra..32v4001E}. The de Sitter precession was detected also by monitoring the heliocentric motion of the Earth-Moon system \citep{1996PhRvD..53.6730W,2009IAU...261.0801W,2018CQGra..35c5015H}, thought of as a giant natural gyroscope,  with the Lunar Laser Ranging (LLR) technique \citep{1994Sci...265..482D}, and in some binary pulsar systems as well \citep{2008Sci...321..104B,Kramer2012}.

I will adopt the scenario by \citet{2013AsBio..13...18H} consisting of a main sequence  star \textcolor{black}{S} orbited at 1 astronomical unit (au) by a gravitationally bound restricted two-body system \textcolor{black}{$\mathcal{S}$} made of a Jupiter-like planet \textcolor{black}{p} and an Earth-mass exomoon \textcolor{black}{s} which, under not too restrictive assumptions, may harbour life. In fact, exomoons may exist also in the habitable zone of M dwarfs \citep{2019ApJ...887..261M,2020A&A...638A..16T}, but, in this case, the analysis would be more involved because of the direct dynamical and tidal effects of the star itself. It will be shown that, for the range of plausible distances allowed to the exomoon in order to be habitable \citep{2013AsBio..13...18H} \textcolor{black}{and a variety of different initial spin-orbit configurations}, \textcolor{black}{$\varepsilon_\mathrm{s}$}  may undergo pN variations \textcolor{black}{$\Delta\varepsilon_\mathrm{s}$ with respect to its initial value $\varepsilon_0^\mathrm{s}$} of tens and even hundreds of degrees  depending on the spin-orbit configuration over $\simeq 0.1-1\,\mathrm{Myr}$.  I will neglect the presence of other planets in the system, so that the ecliptic plane, assumed as reference coordinate $\grf{x,\,y}$ plane, stays \textcolor{black}{essentially} fixed. The orbit of the exomoon around its host planet will be circular, with a size of 5 to 10 planetary radii $R_\mathrm{p}$ \citep{2013AsBio..13...18H}. As far as the exomoon's primary is concerned, I will, first, assume the physical parameters of Jupiter. Then, I will look also at other more massive, larger and more rapidly spinning prototypical gaseous giant planet having the properties of one of those recently characterized in \citet{2020ApJ...905...37B}.
%\textcolor{black}{A similar case is, in fact, present in our Solar System, apart from the system's distance from the Sun: Saturn, whose spin's obliquity to the %ecliptic is \citep{2004AJ....128.2501W} $\varepsilon_\mathrm{p} = 26\grd7$, and its moon Titan, at $20\,R_\mathrm{p}$ from it, whose spin angular momentum ${\bds S}_\mathrm{s}$ is tilted by just %\citep{2008AJ....135.1669S} $\simeq 0.3^\circ$ to its orbital plane which, in turn, lies in the equator of Saturn up to an offset as little as $0.33^\circ$ %\citep[Table\,A.9]{2000ssd..book.....M}. By using the HORIZONS web interface, run by the NASA Jet propulsion Laboratory (JPL), for the inclination and the %node at epoch of the Kronocentric orbit of Titan referred to the International Celestial Reference Frame (ICRF), and \citet{2007CeMDA..98..155S} for the right %ascension (RA) and declination (DEC) at the same epoch of the spin axes of Saturn and Titan, it is possible to infer that the angles $\vartheta$ between $\bds %S,\,\bds L,\,{\bds{S}}_\mathrm{p}$ are as little as $\vartheta_{SS_\mathrm{p}} \simeq 0.6^\circ,\,\vartheta_{SL} \simeq 0.3^\circ,\,\vartheta_{JL} \simeq0.4^\circ$.}}
%\textcolor{black}{Nonetheless, I will allow for more or less marked departures from the ideal alignment condition  by varying, say, their azimuthal angles %and\textcolor{black}{, in another set of runs,} by introducing small offsets in both their spherical angles for all the three angular momenta; if they were %perfectly parallel, no pN spin precessions would occur at all, as it will be shown in Section\,\ref{precess}.}
It should be stressed that, by placing the planet-satellite system at 1 au from the star, the GTR spin precessions considered here are not due to the pN components of the star's gravitational field; moreover, it can be reasonably assumed that the ecliptic plane is not perturbed too much by the post-Keplerian (pK), Newtonian or pN, components of the stellar field which, in principle, are able to induce long-term orbital variations.  Instead, they are induced by the planet's pN field itself; thus, the sources of the de Sitter and Lense-Thirring precessions of the exomoon's spin are the mass $M_\mathrm{p}$ and the spin angular momentum ${\bds S}_\mathrm{p}$ of the Jupiter-type gas giant, respectively \citep{2013grsp.book.....O,2014grav.book.....P}. As such, they are present independently of any peculiar characteristic of the exomoon itself like, e.g., its quadrupole mass moment \textcolor{black}{$J_2^\mathrm{s}$}, Love number \textcolor{black}{$k_2^\mathrm{s}$}, tidal lag time \textcolor{black}{$\Delta t_\mathrm{s}$ \citep{1979M&P....20..301M,2011CeMDA.111..105C,2013ApJ...764...26E}}, etc., which may induce their own long-term changes in the obliquity of its spin. \textcolor{black}{In a further step, I will include also $J_2^\mathrm{s}$ yielding a further direct exomoon's spin precession characterized by a much higher frequency which does not cancel the pN ones superimposing to them.}

The paper is organized as follows. In Section\,\ref{precess}, analytical expressions for the pN rates of change of the satellite's spin axis with respect to a fixed reference plane are derived, and some qualitative features of their solution for $\varepsilon_\mathrm{s}$ are discussed. They are numerically integrated for the case of a Jupiter-like host planet in  Section\,\ref{Jupi}, and for a different gaseous giant body in Section\,\ref{Big} by suitably varying the system's parameter space in both cases. \textcolor{black}{In Section\,\ref{obls}, I deal with the classical spin precession due to the exomoon's own quadrupole mass moment $J_2^\mathrm{s}$, while Section,\ref{other} is devoted to the impact of the distant star and to the obliquity to the planetocentric orbital plane.} Section\,\ref{conclu} summarizes my findings and offer my conclusions.
\section{The de Sitter and Lense-Thirring precessions of the spin's obliquity to the ecliptic plane}\lb{precess}
\subsection{\textcolor{black}{The analytical equations of the pN  precessions}}\lb{analeq}
Let me assume a coordinate system whose reference $\grf{x,\,y}$ plane coincides with the ecliptic plane of the planet-satellite binary. As parameterization of the satellite's spin axis ${\bds{\hat{S}}}_\mathrm{s}$, I adopt
\begin{align}
{\hat{S}}^\mathrm{s}_x \lb{Sx} & = \sin\varepsilon_\mathrm{s}\,\cos\alpha_\mathrm{s}, \\ \nonumber \\
{\hat{S}}^\mathrm{s}_y & = \sin\varepsilon_\mathrm{s}\,\sin\alpha_\mathrm{s}, \\ \nonumber \\
{\hat{S}}^\mathrm{s}_z \lb{Sz}& = \cos\varepsilon_\mathrm{s},
\end{align}
so that $\alpha_\mathrm{s}$ is the spin's azimuthal angle and $\varepsilon_\mathrm{s}$ is its obliquity to the ecliptic: $\varepsilon_\mathrm{s} = 0^\circ$ means that the spin is perpendicular to it. From \rfrs{Sx}{Sz}, it can be straightforwardly obtained
\begin{align}
\dert{\varepsilon_\mathrm{s}}{t} \lb{depsdt}& = -\csc\varepsilon_\mathrm{s}\,\dert{{\hat{S}}^\mathrm{s}_z}{t}, \\ \nonumber \\
\dert{\alpha_\mathrm{s}}{t} \lb{dazdt}& = \csc\varepsilon_\mathrm{s}\,\ton{\cos\alpha_\mathrm{s}\,\dert{{\hat{S}}^\mathrm{s}_y}{t} -\sin\alpha_\mathrm{s}\,\dert{{\hat{S}}^\mathrm{s}_x}{t}  }.
\end{align}

To the pN order, the general relativistic rates of change of $\varepsilon_\mathrm{s},\,\alpha_\mathrm{s}$, averaged over one orbital revolution of the satellite about its parent planet,  can be inferred from the sum of the pN de Sitter and Lense-Thirring averaged precessions of ${\bds{\hat{S}}}_\mathrm{s}$
\eqi
\dert{{\bds{\hat{S}}}_\mathrm{s}}{t} = \ton{ {\mathbf{\Omega}}^\mathrm{s}_\mathrm{dS} + {\mathbf{\Omega}}^\mathrm{s}_\mathrm{LT}}\cross{\bds{\hat{S}}}_\mathrm{s},\lb{zok}
\eqf
where \citep{1975PhRvD..12..329B,2014grav.book.....P}
\begin{align}
{\mathbf{\Omega}}^\mathrm{s}_\mathrm{dS} \lb{dst} & = \rp{3\,\nk\,\mu_\mathrm{p}}{2\,c^2\,a\,\ton{1-e^2}}\bds{\hat{L}}, \\ \nonumber \\
{\mathbf{\Omega}}^\mathrm{s}_\mathrm{LT} \lb{ltr} & =\rp{G\,S_\mathrm{p}}{2\,c^2\,a^3\,\ton{1-e^2}^{3/2}}\qua{{\bds{\hat{S}}}_\mathrm{p} -
3\,\ton{{\bds{\hat{S}}}_\mathrm{p}\bds\cdot\bds{\hat{L}}}\,\bds{\hat{L}}}.
\end{align}
In \rfrs{dst}{ltr}, $c$ is the speed of light in vacuum, $G$ is the Newtonian gravitational constant, $\mu_\mathrm{p}\doteq G\,M_\mathrm{p}$ is the planet's gravitational parameter, $a,\,e$ are the semimajor axis and the eccentricity, respectively, of the satellite's planetocentric orbit, $\nk=\sqrt{\mu_\mathrm{p}/a^3}$ is the Keplerian mean motion,  $\bds{\hat{L}}=\grf{\sin I\,\sin\Omega,\,-\sin I\,\cos\Omega,\,\cos I}$ is the unit vector of the orbital angular momentum, $I$ is the inclination of the satellite's orbital plane to the ecliptic, and $\Omega$ is its longitude of the ascending node.
%, $\bds{\hat{N}}=\grf{\cos\Omega,\,\sin\Omega,\,0}$ is the unit vector directed along the line of the nodes, which is the intersection of the satellite's %orbital plane with the ecliptic one, toward the longitude of the ascending node $\Omega$, and $\bds{\hat{M}}=\grf{-\cos I\,\sin\Omega,\,\cos %I\cos\Omega,\,\sin I}$ is the unit vector directed in the orbital plane such that $\bds{\hat{N}}\cross\bds{\hat{M}} =\bds{\hat{L}}$.
By parameterizing the planet's spin axis ${\bds{\hat{S}}}_\mathrm{p}$ as
\begin{align}
{\hat{S}}^\mathrm{p}_x \lb{Jx} & = \sin\varepsilon_\mathrm{p}\,\cos\alpha_\mathrm{p}, \\ \nonumber \\
{\hat{S}}^\mathrm{p}_y & = \sin\varepsilon_\mathrm{p}\,\sin\alpha_\mathrm{p}, \\ \nonumber \\
{\hat{S}}^\mathrm{p}_z \lb{Jz}& = \cos\varepsilon_\mathrm{p},
\end{align}
where $\varepsilon_\mathrm{p},\,\alpha_\mathrm{p}$ are its obliquity and azimuthal angle, respectively,
from \rfrs{Sx}{Jz}, one finally gets
\begin{align}
\dert{\varepsilon_\mathrm{s}}{t} \nonumber \lb{one} & = -\rp{3\,\nk\,\mu_\mathrm{p}\,\sin I\,\cos\zeta_\mathrm{s} }{2\,c^2\,a\,\ton{1-e^2}} + \\ \nonumber \\
&+ \rp{G\,S_\mathrm{p}}{4\,c^2\,a^3\,\ton{1-e^2}^{3/2}}\,\qua{\sin\varepsilon_\mathrm{p}\ton{\sin\Delta\alpha_\mathrm{sp} - 3\,\sin\chi_\mathrm{sp}} + 3\,\cos\zeta_\mathrm{s} \ton{\cos\varepsilon_\mathrm{p}\,\sin 2I + 2\,\cos^2 I\,\sin\varepsilon_\mathrm{p}\,\sin\zeta_\mathrm{p} }}, \\ \nonumber \\
\dert{\alpha_\mathrm{s}}{t} \nonumber \lb{two} & = \rp{3\,\nk\,\mu_\mathrm{p}\,\ton{\cos I + \cot\varepsilon_\mathrm{s}\,\sin I\,\sin\zeta_\mathrm{s} }}{2\,c^2\,a\,\ton{1-e^2}}
- \\ \nonumber \\
\nonumber & -\rp{G\,S_\mathrm{p}}{8\,c^2\,a^3\,\ton{1-e^2}^{3/2}}\,\grf{\cos\varepsilon_\mathrm{p}\,\ton{2 + 6\,\cos^2 I - 6 \sin^2 I + 6\,\cot\varepsilon_\mathrm{s}\,\sin 2I\,\sin\zeta_\mathrm{s} } + \right. \\ \nonumber \\
&\left. + \sin\varepsilon_\mathrm{p}\,\qua{\ton{\cos\Delta\alpha_\mathrm{sp} + 3\,\cos\chi_\mathrm{sp}}\,\cot\varepsilon_\mathrm{s} -6\,\ton{\sin 2I - \cos 2I\,\cot\varepsilon_\mathrm{s} \sin\zeta_\mathrm{s} }\,\sin\zeta_\mathrm{p} }},
\end{align}
with
\begin{align}
\zeta_\mathrm{s}  \lb{ref1}&\doteq \alpha_\mathrm{s}-\Omega, \\ \nonumber \\
\zeta_\mathrm{p}  \lb{ref3}&\doteq \alpha_\mathrm{p}-\Omega, \\ \nonumber \\
\chi_\mathrm{sp} \lb{ref4} &\doteq \zeta_\mathrm{s}+\zeta_\mathrm{p}, \\ \nonumber \\
\Delta\alpha_\mathrm{sp} \lb{ref2}&\doteq \alpha_\mathrm{s} -\alpha_\mathrm{p}.
\end{align}
\Rfrs{ref1}{ref4} are relative nodes. In terms of the parameterization of \rfrs{Sx}{Sz} and \rfrs{Jx}{Jz}, the azimuthal angle $\Xi$ of $\bds{\hat{L}}$ is related to  $\Omega$ by
\eqi
\Omega=90^\circ + \Xi.
\eqf
If ${\bds{\hat{S}}}_\mathrm{s},\,\bds{\hat{L}},\,{\bds{\hat{S}}}_\mathrm{p}$ are aligned with each other, i.e. for $\varepsilon_\mathrm{s}=I=\varepsilon_\mathrm{p}$ and $\alpha_\mathrm{s}=\Xi=\alpha_\mathrm{p}$, it is
\begin{align}
\zeta_\mathrm{s}  \lb{cond1}& = \zeta_\mathrm{p} = -90^\circ, \\ \nonumber \\
\chi_\mathrm{sp} \lb{cond4}& = -180^\circ,\\ \nonumber \\
\Delta\alpha_\mathrm{sp} & = 0,
\end{align}
so that \rfrs{one}{two} vanish.

In obtaining the averaged rates of \rfrs{one}{two}, it was assumed that $\varepsilon_\mathrm{s},\,\alpha_\mathrm{s}$ stay essentially constant over one satellite's orbital period, and that $I,\,\Omega$ are fixed, i.e. a Keplerian ellipse was used as unperturbed, reference trajectory in the averaging procedure. \textcolor{black}{Such an assumption is justified by the fact that the exomoon's orbital period amounts just to a few days, while the characteristic timescale of its spin precession is of the order of $\simeq 0.1-1\mathrm{Myr}$ (see Sections\,\ref{Jupi}\,to\,\ref{Big}).}
In fact, a long-term modulation is introduced in \rfrs{one}{two} by $I,\,\Omega$ since, actually, they do vary because of a number of classical and general relativistic pK precessions the most important of which are the classical ones due to the planetary oblateness $J_2^\mathrm{p}$ and the pN Lense-Thirring effect caused by the planet's spin ${\bds{S}}_\mathrm{p}$. Their characteristic timescales are much longer than the orbital period by several orders of magnitude since they can be calculated perturbatively by averaging out their Gauss equations over one orbital revolution; the quadrupolar and the Lense-Thirring accelerations \citep{1991ercm.book.....B,SoffelHan19} are, indeed, orders of magnitude smaller than the Newtonian monopole one.
By using \rfrs{Sx}{Sz} and \rfrs{Jx}{Jz}, the classical and relativistic pK averaged rates of change of $I,\,\Omega$ for an arbitrary orientation of the primary's spin axis \citep{1975PhRvD..12..329B,1988NCimB.101..127D,1992PhRvD..45.1840D,2008ApJ...674L..25W,2017EPJC...77..439I}
\begin{align}
\dert I t \lb{iJ2} & = -\rp{3\,\nk\,R_\mathrm{p}^2\,J_2^\mathrm{p}\,\ton{{\bds{\hat{S}}}_\mathrm{p}\bds\cdot\bds{\hat{N}}}\,\ton{{\bds{\hat{S}}}_\mathrm{p}
\bds\cdot\bds{\hat{L}}}}{2\,a^2\,\ton{1-e^2}^2} + \rp{2\,G\,S_\mathrm{p}\,\ton{{\bds{\hat{S}}}_\mathrm{p}\bds\cdot\bds{\hat{N}}}}{c^2\,a^3\,\ton{1-e^2}^{3/2}}, \\ \nonumber \\
\dert\Omega t \lb{NJ2} & = -\rp{3\,\nk\,R_\mathrm{p}^2\,J_2^\mathrm{p}\,\csc I\,\ton{{\bds{\hat{S}}}_\mathrm{p}\bds\cdot\bds{\hat{M}}}\,\ton{{\bds{\hat{S}}}_\mathrm{p}\bds\cdot\textcolor{black}{\bds{\hat{L}}}}}{2\,a^2\,\ton{1-e^2}^2}  + \rp{2\,G\,S_\mathrm{p}\,\csc I\,\ton{{\bds{\hat{S}}}_\mathrm{p}\bds\cdot\bds{\hat{M}}}}{c^2\,a^3\,\ton{1-e^2}^{3/2}},
\end{align}
where $\bds{\hat{N}}=\grf{\cos\Omega,\,\sin\Omega,\,0}$ is the unit vector directed along the line of the nodes, which is the intersection
of the satellite's orbital plane with the ecliptic, toward the longitude of the ascending node,
and $\bds{\hat{M}}=\grf{-\cos I\,\sin\Omega,\,\cos I\,\cos\Omega,\,\sin I}$ is the unit vector directed in the orbital plane such that $\bds{\hat{N}}\bds\times\bds{\hat{M}} = \bds{\hat{L}}$,
can be cast into the form
\begin{align}
\dert I t \lb{IJ2} \nonumber & = \rp{3\,\nk\,R_\mathrm{p}^2\,J_2^\mathrm{p}\,\cos\zeta_\mathrm{p} \,\sin\varepsilon_\mathrm{p}\,\ton{-\cos I\,\cos\varepsilon_\mathrm{p} +\sin I\,\sin\varepsilon_\mathrm{p}\,\sin\zeta_\mathrm{p}   }}{2\,a^2\,\ton{1-e^2}^2}+ \\ \nonumber \\
& +\rp{2\,G\,S_\mathrm{p}\,\sin\varepsilon_\mathrm{p}\,\cos\zeta_\mathrm{p} }{c^2\,a^3\,\ton{1-e^2}^{3/2}}, \\ \nonumber \\
\dert\Omega t \lb{OJ2} \nonumber & = -\rp{3\,\nk\,R_\mathrm{p}^2\,J_2^\mathrm{p}\,\sin I\,\ton{\cos\varepsilon_\mathrm{p} + \cot I\,\sin\varepsilon_\mathrm{p}\,\sin\zeta_\mathrm{p} }\,\ton{\cot I\,\cos\varepsilon_\mathrm{p} - \sin\varepsilon_\mathrm{p}\,\sin\zeta_\mathrm{p} }}{2\,a^2\,\ton{1-e^2}^2} + \\ \nonumber \\
& +\rp{2\,G\,S_\mathrm{p}\,\ton{\cos\varepsilon_\mathrm{p}+\cot I\,\sin\varepsilon_\mathrm{p}\,\sin\zeta_\mathrm{p} }}{c^2\,a^3\,\ton{1-e^2}^{3/2}}.
\end{align}
\rfrs{iJ2}{NJ2} can be expressed in compact, vectorial form as \citep{1975PhRvD..12..329B}
\eqi
\dert{\bds{\hat{L}}}{t} = \ton{{\mathbf{\Omega}}^L_\mathrm{dS} + {\mathbf{\Omega}}^L_\mathrm{LT} + {\mathbf{\Omega}}^L_\mathrm{obl}}\bds\times\bds{\hat{L}},\lb{deLdet}
\eqf
where
\begin{align}
{\mathbf{\Omega}}^L_\mathrm{dS} \lb{bif}&=2\,{\mathbf{\Omega}}^\mathrm{s}_\mathrm{dS}, \\ \nonumber \\
{\mathbf{\Omega}}^L_\mathrm{LT}\lb{bof}&=4\,{\mathbf{\Omega}}^\mathrm{s}_\mathrm{LT},\\ \nonumber \\
{\mathbf{\Omega}}^L_\mathrm{obl} \lb{bloh}&= -\rp{3\,\nk\,J_2^\mathrm{p}\,R_\mathrm{p}^2}{4\,a^2\,\ton{1-e^2}^2}\,\grf{2\ton{{\bds{\hat{S}}}_\mathrm{p}\bds\cdot\bds{\hat{L}}}\,{\bds{\hat{S}}}_\mathrm{p} + \qua{1 - 5\ton{{\bds{\hat{S}}}_\mathrm{p}\bds\cdot\bds{\hat{L}}}^2}\,\bds{\hat{L}}  }.
\end{align}
Note that \rfrs{bif}{bloh} strictly hold in the test particle limit of the full two-body expressions by \citet{1975PhRvD..12..329B}.
\rfrs{IJ2}{OJ2} vanish for \rfrs{cond1}{cond4}. Thus, also the orbital angular momentum stay fixed in space if ${\bds{\hat{S}}}_\mathrm{s},\,\bds{\hat{L}},\,{\bds{\hat{S}}}_\mathrm{p}$ are aligned, as it can straightforwardly be inferred from \rfrs{dst}{ltr}, \rfr{bloh} and \rfr{deLdet}.

\rfrs{one}{two} and \rfrs{IJ2}{OJ2} represent a system of nonlinear first order differential equations  for the four unknowns $\varepsilon_\mathrm{s},\,\alpha_\mathrm{s},\,I,\,\Omega$ to be simultaneously integrated;  $\varepsilon_\mathrm{p},\,\alpha_\mathrm{p}$ are assumed to be constant.
\subsection{Some qualitative features of the pN obliquity precessions}\lb{referee}
In order to grasp some essential features of the numerically integrated time series displayed in Sections\,\ref{Jupi}\,to\,\ref{Big}, I will make some considerations about \rfrs{one}{two} and \rfrs{IJ2}{OJ2}.

Let me define
\begin{align}
\nu_\mathrm{dS} \lb{nudS}&\doteq \rp{3\,\nk\,\mu_\mathrm{p}}{2\,c^2\,a\,\ton{1-e^2}}, \\ \nonumber \\
\nu_\mathrm{LT} \lb{nuLT}& \doteq \rp{G\,S_\mathrm{p}}{c^2\,a^3\,\ton{1-e^2}^{3/2}},\\ \nonumber\\
\nu_\mathrm{obl} &\doteq \rp{3\,\nk\,J_2^\mathrm{p}\,R_\mathrm{p}^2}{2\,a^2\,\ton{1-e^2}^2},
\end{align}
so that
\begin{align}
\Lambda_\mathrm{dS}^\mathrm{LT} \lb{Lam}&\doteq \rp{\nu_\mathrm{LT}}{\nu_\mathrm{dS}}=\rp{2\,S_\mathrm{p}}{3\,\sqrt{M_\mathrm{p}^3\,G\,a\,\ton{1-e^2}}},\\ \nonumber \\
\Delta^\mathrm{LT}_\mathrm{obl}&\doteq \rp{2\,\nu_\mathrm{LT}}{\nu_\mathrm{obl}}=\rp{4\,G\,S_\mathrm{p}\,\sqrt{1-e^2}}{3\,c^2\,a\,\nk\,R_\mathrm{p}^2\,J_2^\mathrm{p}}, \\ \nonumber \\
\Gamma_\mathrm{dS}^\mathrm{obl}&\doteq \rp{\nu_\mathrm{obl}}{\nu_\mathrm{dS}}=\rp{c^2\,J_2^\mathrm{p}\,R_\mathrm{p}^2}{\mu_\mathrm{p}\,a\,\ton{1-e^2}}.\lb{frz}
\end{align}
For a circular orbit at $5\,R_\mathrm{p}\leq a\leq 10\,R_\mathrm{p}$ from a Jupiter-like host planet, it is
\begin{align}
\Lambda_\mathrm{dS}^\mathrm{LT}\lb{LampN}&\simeq 0.03, \\ \nonumber \\
\Delta^\mathrm{LT}_\mathrm{obl}&\simeq 5\times 10^{-7}, \\ \nonumber \\
\Gamma_\mathrm{dS}^\mathrm{obl}&\simeq 10^5.
\end{align}
It implies that, in the considered scenario \textcolor{black}{and over timescales shorter than the gravitomagnetic characteristic ones,} the Lense-Thirring spin and orbital precessions can generally be neglected with respect to the de Sitter and quadrupolar ones, respectively, apart from some particular spin and orbital configurations.
Also in such an approximated case, it is not possible to obtain analytical solutions from the simultaneous integration of \rfrs{one}{two} and \rfrs{IJ2}{OJ2} for an arbitrary spin-orbit configuration. Nonetheless, some qualitative features, which will allow to understand certain aspects of the numerically integrated time series in Section\,\ref{Big} with respect to those in Section\,\ref{Jupi}, can still be inferred.
\Rfrs{one}{two} can be combined obtaining
\eqi
\dert{\varepsilon_\mathrm{s}}{\alpha_\mathrm{s}} \lb{noia1}\simeq -\rp{\tan I\,\cos\zeta_\mathrm{s} }{1+\tan I\,\cot\varepsilon_\mathrm{s}\,\sin\zeta_\mathrm{s} },
\eqf
where
\eqi
\dert{\alpha_\mathrm{s}}{t} \lb{noia2}\simeq \nu_\mathrm{dS}\,\cos I\,\ton{1+\tan I\,\cot\varepsilon_\mathrm{s}\,\sin\zeta_\mathrm{s} }.
\eqf
\Rfr{noia1} implies that, whether $I$ and $\Omega$ vary or not, the amplitude of $\varepsilon(\alpha_\mathrm{s})$ is independent of either the exomoon's distance from the planet and the physical parameters of the latter up to the order of $\mathcal{O}\ton{\Lambda_\mathrm{dS}^\mathrm{LT}}$, depending only on the initial spin-orbit configuration. Furthermore, \rfr{noia2} tells that, however complicated the dependence of $\alpha_\mathrm{s}$ on $t$ may be, the characteristic time scale of the resulting time series of $\varepsilon_\mathrm{s}\ton{t}$, which is generally not a simple harmonic function, is determined by \rfr{nudS}. Such features can be explicitly inferred in some particular cases like, e.g., in a purely de Sitter scenario. According to \rfrs{IJ2}{OJ2},  it is $I=I_0,\,\Omega=\Omega_0$ for $J_2^\mathrm{p} =0,\,S_\mathrm{p}=0$. Then, by analytically solving \rfr{zok} with \rfr{dst} only,  it is possible to obtain
\begin{align}
{\hat{S}}^\mathrm{s}_z\ton{t} \nonumber &=\cos\varepsilon_\mathrm{s}\ton{t} = \cos^2 I_0\,\cos \varepsilon_\mathrm{s}^0 + \cos \varepsilon_\mathrm{s}^0\,\cos\ton{\nu_\mathrm{dS}\,t}\,\sin^2 I_0 + \sin \varepsilon_\mathrm{s}^0\,\qua{\cos\zeta_\mathrm{s}^0\,\sin I_0\,\sin\ton{\nu_\mathrm{dS}\,t} -\right.\\ \nonumber \\
&-\left.\sin 2 I_0\,\sin^2\ton{\rp{\nu_\mathrm{dS}}{2}\,t}\,\sin\zeta_\mathrm{s}^0},
\end{align}
where $\zeta_\mathrm{s}^0\doteq \alpha_\mathrm{s}^0-\Omega_0$.
\section{The case of a Jupiter-like parent planet}\lb{Jupi}
I  numerically integrated \rfrs{one}{two} along with \rfrs{IJ2}{OJ2} over $1\,\mathrm{Myr}$ for a fictitious exomoon circling a planet with the same physical parameters of Jupiter along a circular orbit at a few radii from it, as per \citet{2013AsBio..13...18H}. About the initial conditions, I, first, looked at different mutual orientations of ${{\bds{\hat{S}}}_\mathrm{s}},\,{\bds{\hat{L}}},\,{\bds{\hat{S}}}_\mathrm{p}$ for a fixed satellite's planetocentric distance $a=5\,R_\mathrm{p}$\textcolor{black}{; restricting only to, say, a Jupiter-Io scenario would not be justified by the different  possible formation mechanisms of exomoons, by the variety of spin-orbit configurations  in several star-exoplanet systems discovered so far \citep{2016ApJ...819...85C}, and by the fact that, until now, no exomoons have yet been unquestionably detected}. Then, in Figure\,\ref{fig5}, I varied $a$ from $5\,R_\mathrm{p}$ to $10\,R_\mathrm{p}$  for a given spin-orbit configuration.

\begin{figure}[H]
\centering
\centerline{
\vbox{
\begin{tabular}{c}
\epsfxsize= 16 cm\epsfbox{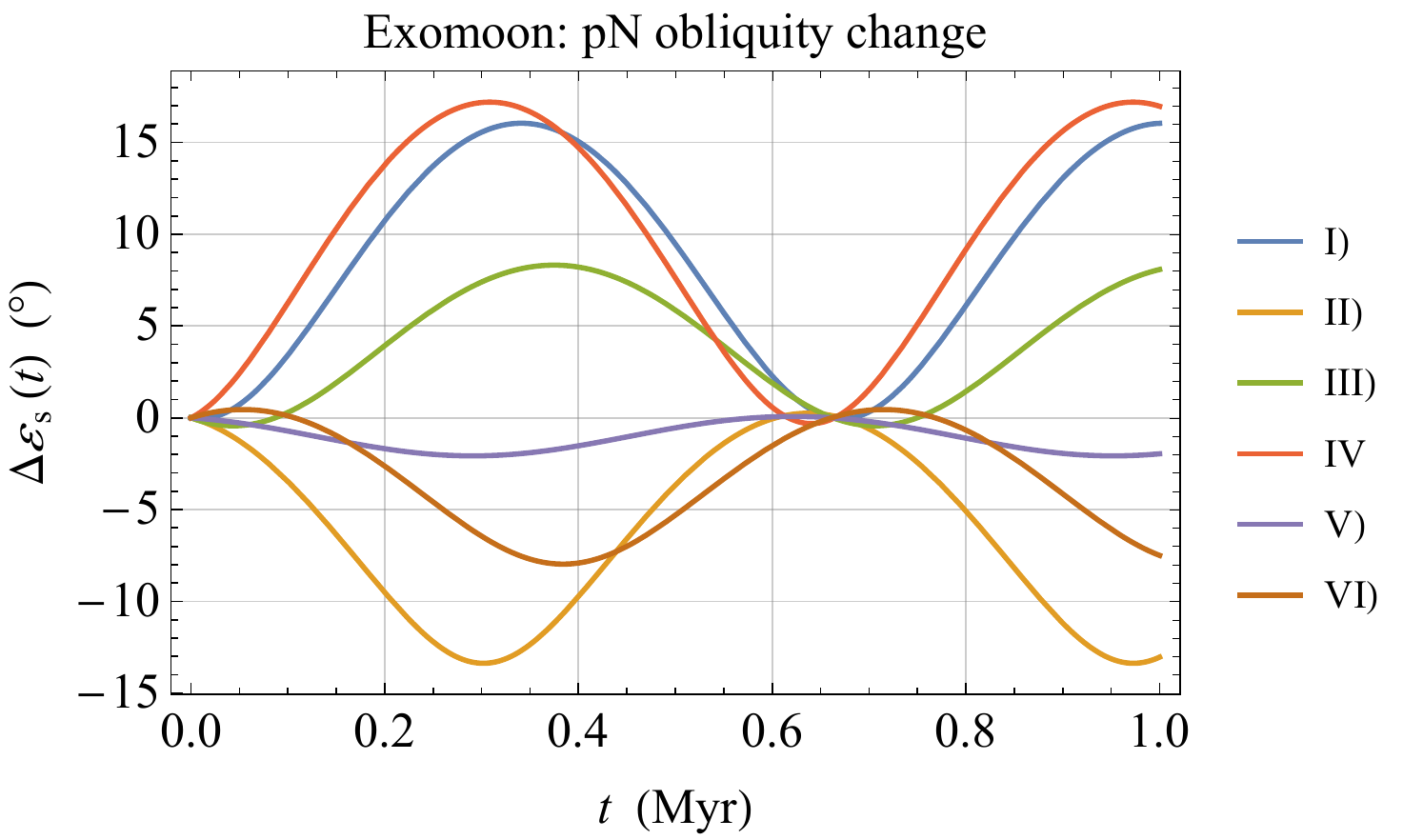}\\
\end{tabular}
}
}
\caption{
Numerically produced time series $\Delta\varepsilon_\mathrm{s}\ton{t} = \varepsilon_\mathrm{s}\ton{t}-\varepsilon_\mathrm{s}^0$, in $^\circ$, of the general relativistic pN variation of the obliquity $\varepsilon_\mathrm{s}$ to the ecliptic plane of a putative exomoon orbiting a gaseous giant planet with the same physical properties of Jupiter. They were obtained by simultaneously integrating  the orbit-averaged \rfrs{one}{two} and \rfrs{IJ2}{OJ2}  for the rates of change of $\varepsilon_\mathrm{s},\,\alpha_\mathrm{s},\,\Omega,\,I$ over $1\,\mathrm{Myr}$.  The initial conditions, corresponding to a close alignment of $\bds{\hat{L}},\,{\bds{\hat{S}}}_\mathrm{p},\,{\bds{\hat{S}}}_\mathrm{s}$, are listed in Table\,\ref{tab1}, which summarizes the main features of the signatures as well.
}\label{fig1}
\end{figure}

\textcolor{black}{The case of a close, although not perfect, mutual alignment of $\bds{\hat{L}},\,{\bds{\hat{S}}}_\mathrm{p},\,{\bds{\hat{S}}}_\mathrm{s}$ is shown in Figure\,\ref{fig1}, based on the initial conditions of Table\,\ref{tab1}. I allowed for  offsets of a few degrees among the spherical angles of the three angular momenta with respect to, say, $\grf{\theta}=23\grd44,\,\grf{\phi}=150^\circ$, where $\grf{\theta}\doteq \varepsilon_\mathrm{s}^0,\,I_0,\,\varepsilon_\mathrm{p}$ and $\grf{\phi}=\alpha_\mathrm{s}^0,\,\Xi_0,\,\alpha_\mathrm{p}$.
The resulting ranges $\varepsilon_\mathrm{s}^\mathrm{max}-\varepsilon_\mathrm{s}^\mathrm{min}$ of the time series of Figure\,\ref{fig1} are still potentially significant for life since their sizes are of the order of $\simeq 3^\circ-17^\circ$ over characteristic timescales as short as $0.7\,\mathrm{Myr}$, as resumed in Table\,\ref{tab1}.
}
\begin{table*}
\caption{Initial conditions used in Figure\,\ref{fig1}. Each row corresponds to the plotted times series with the same roman numeral in the legend of Figure\,\ref{fig1}. Recall that the azimuthal angle of $\bds{\hat{L}}$ is $\Xi=\Omega-90^\circ$. $T$ is the characteristic timescale, and $\varepsilon_\mathrm{s}^\mathrm{max}-\varepsilon_\mathrm{s}^\mathrm{min}$ is the full range of variation of the exomoon's obliquity to the ecliptic.}
\label{tab1}
\centering
\begin{tabular}{ccccccccccc}
\noalign{\smallskip}
\hline
  & $a$ ($R_\mathrm{p}$) & $e$ & $I_0$ ($^\circ$) & $\Omega_0$ ($^\circ$) & $\varepsilon_\mathrm{p}$ ($^\circ$) & $\alpha_\mathrm{p}$ ($^\circ$) & $\varepsilon_\mathrm{s}^0$ ($^\circ$) & $\alpha_\mathrm{s}^0$ ($^\circ$) & $T$ (Myr) & $\varepsilon_\mathrm{s}^\mathrm{max}-\varepsilon_\mathrm{s}^\mathrm{min}$ ($^\circ$)\\
\hline
I) & $5$ & $0.0$ & $25\grd44$ & $233$ & $29\grd44$ & $147$ & $21\grd44$ & $149$ & $0.7$ & $16$  \\
II) & $5$ & $0.0$ & $28\grd44$ & $243$ & $18\grd44$ & $146$ & $25$ & $151$ & $0.7$ & $13$  \\
III) & $5$ & $0.0$ & $22\grd44$ & $248$ & $24\grd44$ & $148$ & $20.50$ & $153$ & $0.7$ & $10$  \\
IV) & $5$ & $0.0$ & $21\grd7$ & $242$ & $27\grd44$ & $156$ & $19$ & $150$ & $0.7$ & $17$  \\
V) & $5$ & $0.0$ & $20\grd44$ & $240$ & $22\grd44$ & $153$ & $23\grd44$ & $154$ & $0.7$ & $3$  \\
VI) & $5$ & $0.0$ & $19\grd44$ & $238$ & $20\grd44$ & $152$ & $24\grd20$ & $147$ & $0.7$ & $9$  \\
\hline
\end{tabular}
\end{table*}

\begin{figure}[H]
\centering
\centerline{
\vbox{
\begin{tabular}{c}
\epsfxsize= 16 cm\epsfbox{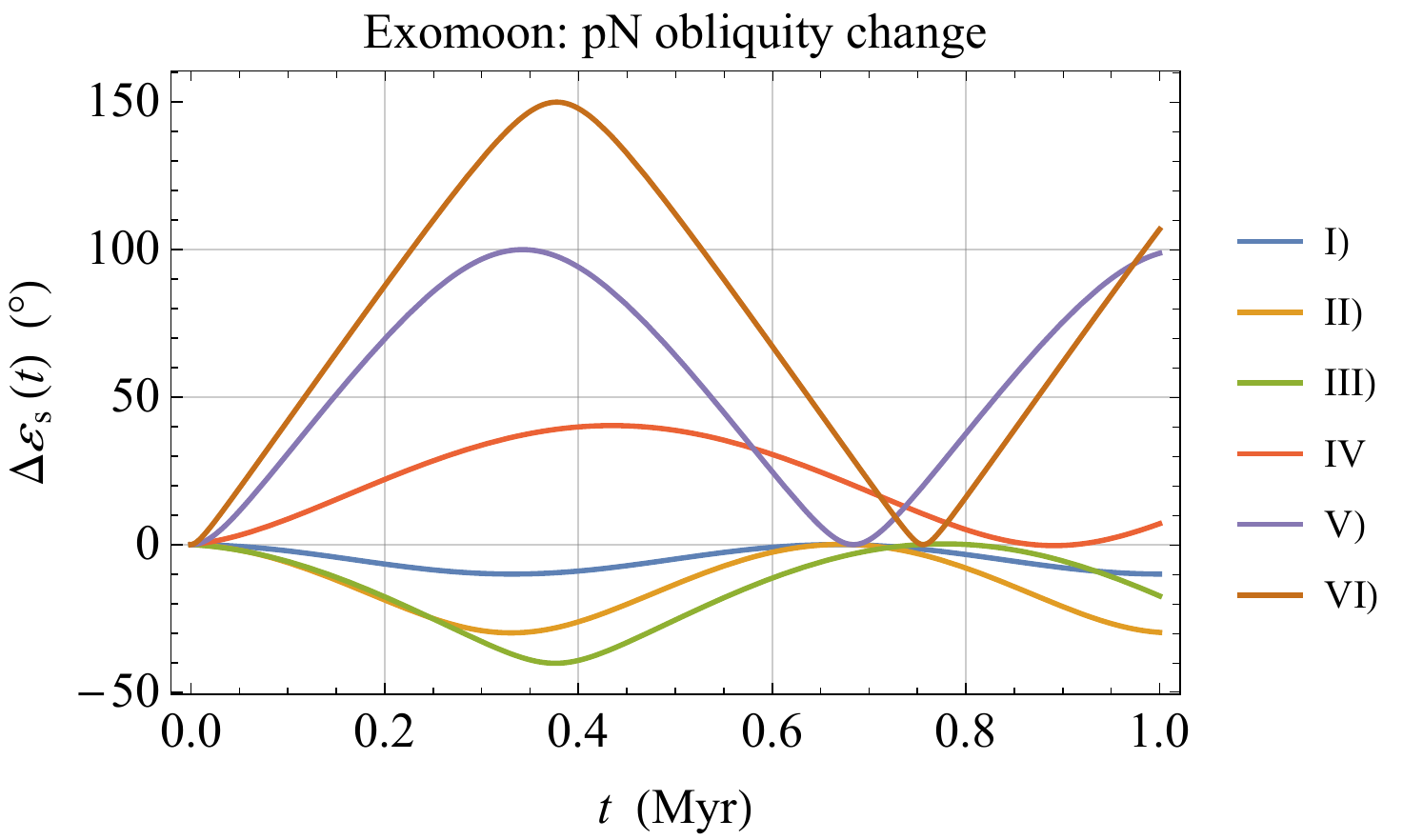}\\
\end{tabular}
}
}
\caption{
Numerically produced time series $\Delta\varepsilon_\mathrm{s}\ton{t} = \varepsilon_\mathrm{s}\ton{t}-\varepsilon_\mathrm{s}^0$, in $^\circ$, of the general relativistic pN variation of the obliquity $\varepsilon_\mathrm{s}$ to the ecliptic plane of a putative exomoon orbiting a gaseous giant planet with the same physical properties of Jupiter. They were obtained by simultaneously integrating  the orbit-averaged \rfrs{one}{two} and \rfrs{IJ2}{OJ2}  for the rates of change of $\varepsilon_\mathrm{s},\,\alpha_\mathrm{s},\,\Omega,\,I$ over $1\,\mathrm{Myr}$.  The initial conditions, corresponding to $\bds{\hat{L}},\,{\bds{\hat{S}}}_\mathrm{p},\,{\bds{\hat{S}}}_\mathrm{s}$ lying almost in the same azimuthal plane but tilted differently from each other, are listed in Table\,\ref{tab2}, which summarizes the main features of the signatures as well.
}\label{fig2}
\end{figure}

Table\,\ref{tab2} and Figure\,\ref{fig2} deal with the case in which the three angular momenta initially share almost the same azimuthal plane, being tilted differently to the ecliptic. It can be noted that $\varepsilon_\mathrm{s}^\mathrm{max}-\varepsilon_\mathrm{s}^\mathrm{min}$ ranges from $10^\circ$ to $150^\circ$.
\begin{table*}
\caption{Initial conditions used in Figure\,\ref{fig2}. Each row corresponds to the plotted times series with the same roman numeral in the legend of Figure\,\ref{fig2}. Recall that the azimuthal angle of $\bds{\hat{L}}$ is $\Xi=\Omega-90^\circ$. $T$ is the characteristic timescale, and $\varepsilon_\mathrm{s}^\mathrm{max}-\varepsilon_\mathrm{s}^\mathrm{min}$ is the full range of variation of the exomoon's obliquity to the ecliptic.}
\label{tab2}
\centering
\begin{tabular}{ccccccccccc}
\noalign{\smallskip}
\hline
  & $a$ ($R_\mathrm{p}$) & $e$ & $I_0$ ($^\circ$) & $\Omega_0$ ($^\circ$) & $\varepsilon_\mathrm{p}$ ($^\circ$) & $\alpha_\mathrm{p}$ ($^\circ$) & $\varepsilon_\mathrm{s}^0$ ($^\circ$) & $\alpha_\mathrm{s}^0$ ($^\circ$) & $T$ (Myr) & $\varepsilon_\mathrm{s}^\mathrm{max}-\varepsilon_\mathrm{s}^\mathrm{min}$ ($^\circ$)\\
\hline
I) & $5$ & $0.0$ & $15$ & $233$ & $5$ & $147$ & $80$ & $149$ & $0.7$ & $10$  \\
II) & $5$ & $0.0$ & $30$ & $243$ & $15$ & $146$ & $65$ & $151$ & $0.7$ & $30$  \\
III) & $5$ & $0.0$ & $65$ & $248$ & $30$ & $148$ & $50$ & $153$ & $0.7$ & $40$  \\
IV) & $5$ & $0.0$ & $5$ & $242$ & $50$ & $156$ & $30$ & $150$ & $0.7$ & $40$  \\
V) & $5$ & $0.0$ & $80$ & $240$ & $65$ & $153$ & $15$ & $154$ & $0.7$ & $100$  \\
VI) & $5$ & $0.0$ & $50$ & $238$ & $80$ & $152$ & $5$ & $147$ & $0.7$ & $150$  \\
\hline
\end{tabular}
\end{table*}

\begin{figure}[H]
\centering
\centerline{
\vbox{
\begin{tabular}{c}
\epsfxsize= 16 cm\epsfbox{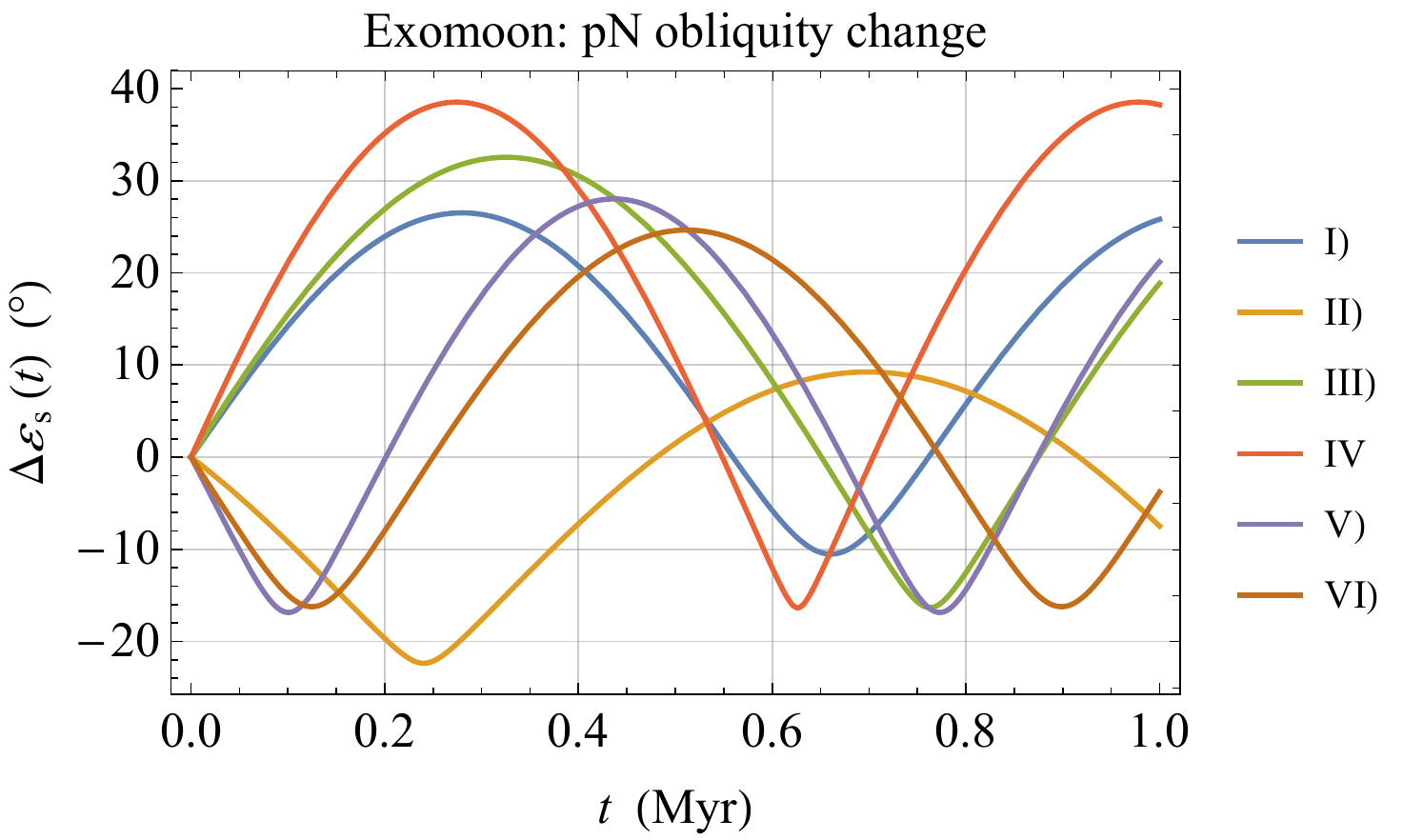}\\
\end{tabular}
}
}
\caption{
Numerically produced time series $\Delta\varepsilon_\mathrm{s}\ton{t} = \varepsilon_\mathrm{s}\ton{t}-\varepsilon_\mathrm{s}^0$, in $^\circ$, of the general relativistic pN variation of the obliquity $\varepsilon_\mathrm{s}$ to the ecliptic plane of a putative exomoon orbiting a gaseous giant planet with the same physical properties of Jupiter. They were obtained by simultaneously integrating  the orbit-averaged \rfrs{one}{two} and \rfrs{IJ2}{OJ2}  for the rates of change of $\varepsilon_\mathrm{s},\,\alpha_\mathrm{s},\,\Omega,\,I$ over $1\,\mathrm{Myr}$.  The initial conditions, corresponding to $\bds{\hat{L}},\,{\bds{\hat{S}}}_\mathrm{p},\,{\bds{\hat{S}}}_\mathrm{s}$ tilted almost identically to the ecliptic but located in different azimuthal planes, are listed in Table\,\ref{tab3}, which summarizes the main features of the signatures as well.
}\label{fig3}
\end{figure}

Table\,\ref{tab3} and Figure\,\ref{fig3} refer to  $\bds{\hat{L}},\,{\bds{\hat{S}}}_\mathrm{p},\,{\bds{\hat{S}}}_\mathrm{s}$  sharing almost the same tilt to the ecliptic and displaced in different azimuthal planes. In this case, the range of values for $\Delta\varepsilon_\mathrm{s}$ is narrower than in Figures\,\ref{fig1}\,to\,\ref{fig2}, amounting to $36^\circ\leq\varepsilon_\mathrm{s}^\mathrm{max}-\varepsilon_\mathrm{s}^\mathrm{min}\leq 54^\circ$. Nonetheless, it still remains likely significative for the exomoon's habitability.
\begin{table*}
\caption{Initial conditions used in Figure\,\ref{fig3}. Each row corresponds to the plotted times series with the same roman numeral in the legend of Figure\,\ref{fig3}. Recall that the azimuthal angle of $\bds{\hat{L}}$ is $\Xi=\Omega-90^\circ$. $T$ is the characteristic timescale, and $\varepsilon_\mathrm{s}^\mathrm{max}-\varepsilon_\mathrm{s}^\mathrm{min}$ is the full range of variation of the exomoon's obliquity to the ecliptic. }
\label{tab3}
\centering
\begin{tabular}{ccccccccccc}
\noalign{\smallskip}
\hline
  & $a$ ($R_\mathrm{p}$) & $e$ & $I_0$ ($^\circ$) & $\Omega_0$ ($^\circ$) & $\varepsilon_\mathrm{p}$ ($^\circ$) & $\alpha_\mathrm{p}$ ($^\circ$) & $\varepsilon_\mathrm{s}^0$ ($^\circ$) & $\alpha_\mathrm{s}^0$ ($^\circ$) & $T$ (Myr) & $\varepsilon_\mathrm{s}^\mathrm{max}-\varepsilon_\mathrm{s}^\mathrm{min}$ ($^\circ$)\\
\hline
I) & $5$ & $0.0$ & $25\grd44$ & $90$ & $29\grd44$ & $290$ & $21\grd44$ & $250$ & $0.7$ & $37$  \\
II) & $5$ & $0.0$ & $28\grd44$ & $130$ & $18\grd44$ & $250$ & $25$ & $290$ & $0.7$ & $36$  \\
III) & $5$ & $0.0$ & $22\grd44$ & $170$ & $24\grd44$ & $210$ & $20.50$ & $130$ & $0.7$ & $48$  \\
IV) & $5$ & $0.0$ & $21\grd7$ & $210$ & $27\grd44$ & $170$ & $19$ & $90$ & $0.7$ & $54$  \\
V) & $5$ & $0.0$ & $20\grd44$ & $250$ & $22\grd44$ & $130$ & $23\grd44$ & $210$ & $0.7$ & $44$  \\
VI) & $5$ & $0.0$ & $19\grd44$ & $290$ & $20\grd44$ & $90$ & $24\grd20$ & $170$ & $0.7$ & $40$  \\
\hline
\end{tabular}
\end{table*}

\begin{figure}[H]
\centering
\centerline{
\vbox{
\begin{tabular}{c}
\epsfxsize= 16 cm\epsfbox{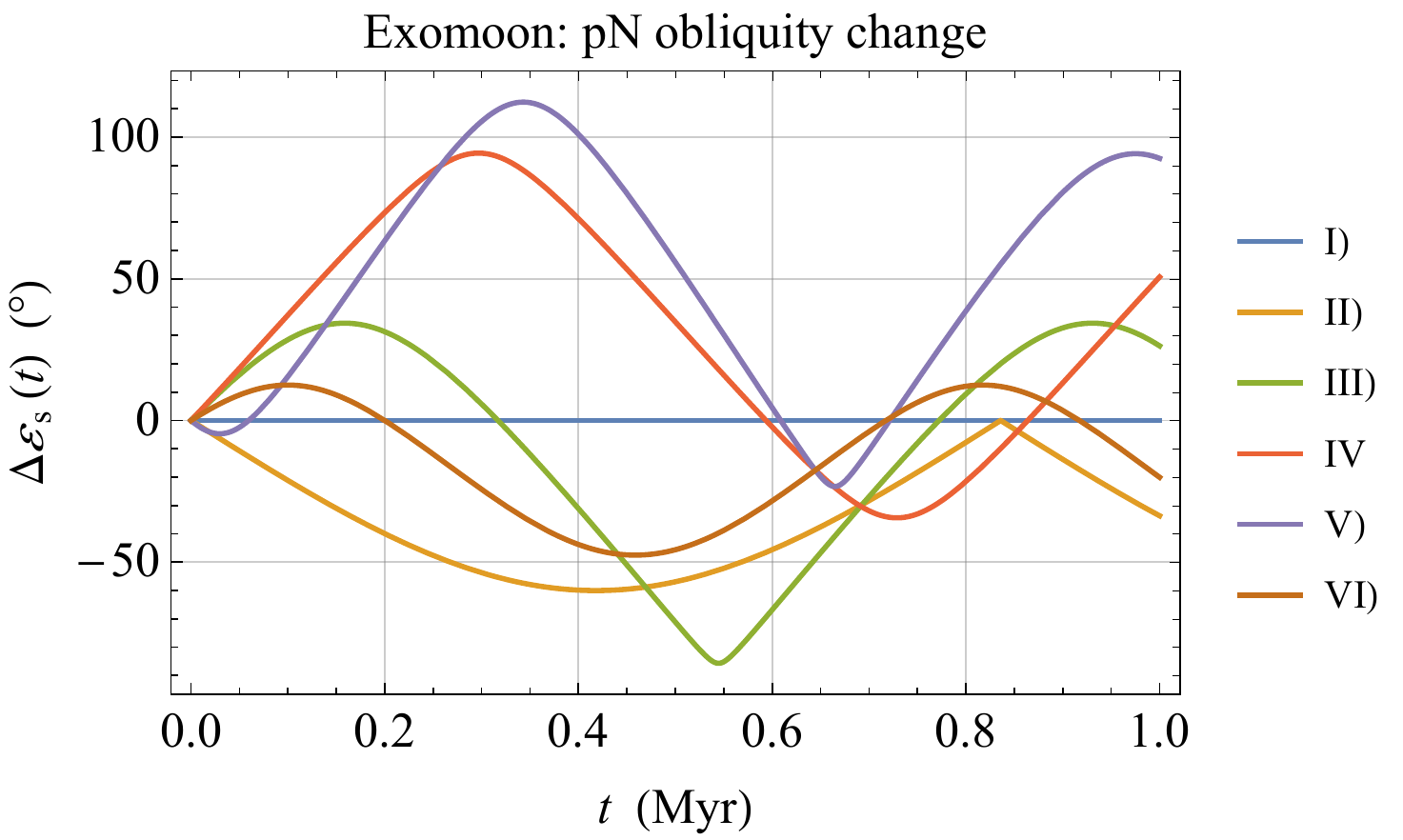}\\
\end{tabular}
}
}
\caption{
Numerically produced time series $\Delta\varepsilon_\mathrm{s}\ton{t} = \varepsilon_\mathrm{s}\ton{t}-\varepsilon_\mathrm{s}^0$, in $^\circ$, of the general relativistic pN variation of the obliquity $\varepsilon_\mathrm{s}$ to the ecliptic plane of a putative exomoon orbiting a gaseous giant planet with the same physical properties of Jupiter. They were obtained by simultaneously integrating  the orbit-averaged \rfrs{one}{two} and \rfrs{IJ2}{OJ2}  for the rates of change of $\varepsilon_\mathrm{s},\,\alpha_\mathrm{s},\,\Omega,\,I$ over $1\,\mathrm{Myr}$.  The initial conditions, corresponding to $\bds{\hat{L}},\,{\bds{\hat{S}}}_\mathrm{p},\,{\bds{\hat{S}}}_\mathrm{s}$ arbitrarily oriented, are listed in Table\,\ref{tab4}, which summarizes the main features of the signatures as well.
}\label{fig4}
\end{figure}

\begin{figure}[H]
\centering
\centerline{
\vbox{
\begin{tabular}{c}
\epsfxsize= 16 cm\epsfbox{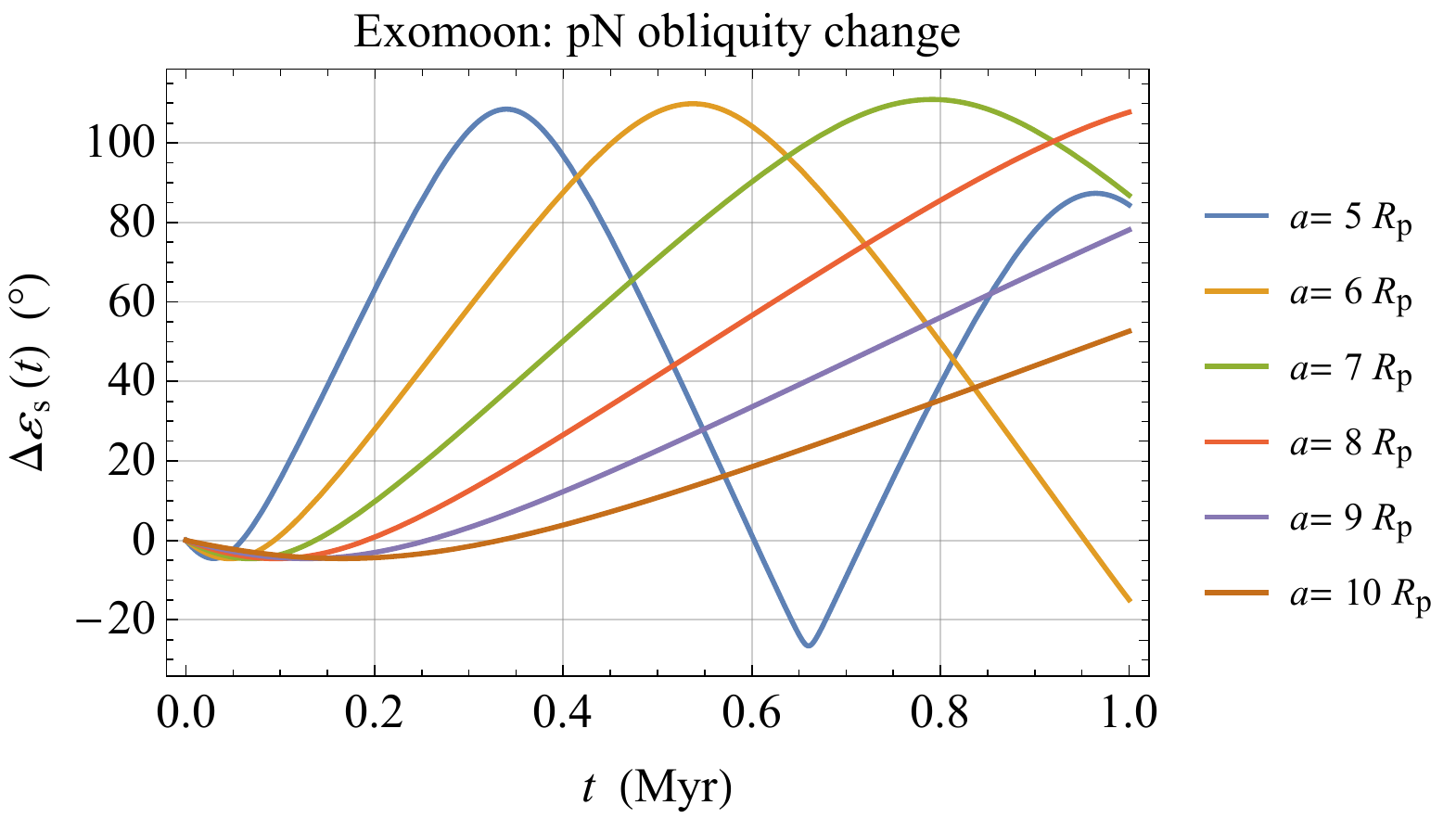}\\
\end{tabular}
}
}
\caption{
Numerically produced time series $\Delta\varepsilon_\mathrm{s}\ton{t} = \varepsilon_\mathrm{s}\ton{t}-\varepsilon_\mathrm{s}^0$, in $^\circ$, of the general relativistic pN variation of the obliquity $\varepsilon_\mathrm{s}$ to the ecliptic plane of a putative exomoon orbiting a gaseous giant planet with the same physical properties of Jupiter. They were obtained by simultaneously integrating  the orbit-averaged \rfrs{one}{two} and \rfrs{IJ2}{OJ2}  for the rates of change of $\varepsilon_\mathrm{s},\,\alpha_\mathrm{s},\,\Omega,\,I$ over $1\,\mathrm{Myr}$.  The initial spin-orbit configuration, common to all the runs, is listed in Table\,\ref{tab5}.
}\label{fig5}
\end{figure}

The case for an arbitrary mutual orientation of the three angular momenta is displayed by Table\,\ref{tab4} and Figure\,\ref{fig4}. The  range of values for the satellite's obliquity to the ecliptic is $60^\circ-132^\circ$.
\begin{table*}
\caption{Initial conditions used in Figure\,\ref{fig4}. Each row corresponds to the plotted times series with the same roman numeral in the legend of Figure\,\ref{fig4}. Recall that the azimuthal angle of $\bds{\hat{L}}$ is $\Xi=\Omega-90^\circ$. $T$ is the characteristic timescale, and $\varepsilon_\mathrm{s}^\mathrm{max}-\varepsilon_\mathrm{s}^\mathrm{min}$ is the full range of variation of the exomoon's obliquity to the ecliptic.}
\label{tab4}
\centering
\begin{tabular}{ccccccccccc}
\noalign{\smallskip}
\hline
  & $a$ ($R_\mathrm{p}$) & $e$ & $I_0$ ($^\circ$) & $\Omega_0$ ($^\circ$) & $\varepsilon_\mathrm{p}$ ($^\circ$) & $\alpha_\mathrm{p}$ ($^\circ$) & $\varepsilon_\mathrm{s}^0$ ($^\circ$) & $\alpha_\mathrm{s}^0$ ($^\circ$) & $T$ (Myr) & $\varepsilon_\mathrm{s}^\mathrm{max}-\varepsilon_\mathrm{s}^\mathrm{min}$ ($^\circ$)\\
\hline
I)   & $5$ & $0.0$ & $60$ & $0$  & $180$ & $300$ & $150$ & $120$ & $0.7$ & $0.0012$  \\
II)  & $5$ & $0.0$ & $30$ & $60$ & $150$ & $240$ & $180$ & $300$ & $0.7$ & $60$  \\
III) & $5$ & $0.0$ & $150$ & $120$ & $60$ & $180$ & $90$ & $240$ & $0.7$ & $110$  \\
IV)  & $5$ & $0.0$ & $120$ & $180$ & $90$ & $120$ & $60$ & $0$  & $0.7$ & $120$  \\
V)   & $5$ & $0.0$ & $90$ & $240$ & $120$ & $60$ & $30$ & $180$ & $0.7$ & $132$  \\
VI)  & $5$ & $0.0$ & $180$ & $300$ & $30$ & $0$  & $120$ & $60$ & $0.7$ & $60$  \\
\hline
\end{tabular}
\end{table*}
In Figure\,\ref{fig5}, obtained for the initial spin-orbit configuration of Table\,\ref{tab5}, I varied the satellite's planetocentric distance from $5\,R_\mathrm{p}$ to $10\,R_\mathrm{p}$. As expected from the qualitative analysis of Section\,\ref{referee}, each time series retains essentially the same maximum range of values $\varepsilon_\mathrm{s}^\mathrm{max}-\varepsilon_\mathrm{s}^\mathrm{min}\simeq 180^\circ$, while the characteristic timescales of their
%complicated, non-harmonic
temporal patterns, determined by \rfr{nudS}, increase with $a$.
\begin{table*}
\caption{Initial spin-orbit configuration used in Figure\,\ref{fig5}. }
\label{tab5}
\centering
\begin{tabular}{cccccc}
\noalign{\smallskip}
\hline
  $I_0$ ($^\circ$) & $\Omega_0$ ($^\circ$) & $\varepsilon_\mathrm{p}$ ($^\circ$) & $\alpha_\mathrm{p}$ ($^\circ$) & $\varepsilon_\mathrm{s}^0$ ($^\circ$) & $\alpha_\mathrm{s}^0$ ($^\circ$)\\
\hline
 $90$ & $240$  & $120$ & $60$ & $30$ & $180$ \\
\hline
\end{tabular}
\end{table*}

I checked the validity of Figures\,\ref{fig1}\,to\,\ref{fig5}, based on the orbit-averaged \rfrs{one}{two} and \rfrs{IJ2}{OJ2}, by numerically integrating the precessional equations of both the satellite's spin and orbital angular momenta in the vectorial form given by \citet{1975PhRvD..12..329B} over the same time spans of the previous runs. In particular, by neglecting all the contributions from the exomoon's own spin ${\bds S}_\mathrm{s}$ and quadrupole mass moment $J_2^\mathrm{s}$, I used \rfrs{dst}{ltr} for the satellite's spin, and \rfrs{deLdet}{bloh} for the orbital angular momentum.
I adopted the same orbital and physical parameters of the previous integrations along with the same initial conditions.
Then, I computed the time series for $\varepsilon_\mathrm{s}\ton{t}$ from the solution for ${\hat{S}}^\mathrm{s}_z\ton{t}$
as $\varepsilon_\mathrm{s}\ton{t} = \arccos{\hat{S}}^\mathrm{s}_z\ton{t}$ by obtaining curves indistinguishable from those in  Figures\,\ref{fig1}\,to\,\ref{fig5} since they differ, at most, by much less than $1^\circ$.

The obliquity $\varepsilon_\mathrm{s}$  can, in principle, vary also because of torques of classical origin which, however, depend on the peculiar characteristics of the satellite like its own oblateness $J_2^\mathrm{s}$ of both tidal and centrifugal origin \citep{2009ApJ...698.1778R}. Instead, the pN effects previously investigated are due only to the spacetime itself deformed by the mass-energy currents of the parent planet. \textcolor{black}{In Section\,\ref{obls}, I will look at the impact of $J_2^\mathrm{s}$ as well.}

Variations of the latitudinal insolation received by the exomoon from the host star due to changes in $\varepsilon_\mathrm{s}$ so large as those exhibited by Figures\,\ref{fig1}\,to\,\ref{fig5}
%, which would affect also the amount of radiation received by the planet as well,
can certainly have a sensible impact on its habitability, representing a novel element which should be taken into account in future studies on the capability of such worlds to sustain life.
\section{The case of faster spinning, larger and more massive gaseous giant planets}\lb{Big}
Until now, I limited myself to the case of a host planet with the same Jovian physical parameters. It is important to look also at other possible gaseous giants with different fundamental characteristics with respect to Jupiter.
Recently, for some of them orbiting at a few tens or hundreds of astronomical units from their parent stars, it was possible to determine some key parameters like the mass $M_\mathrm{p}$, the equatorial radius $R_\mathrm{p}$ and the spinning period $P_\mathrm{p}$ \citep{2020ApJ...905...37B}. In order to assess the pN effects on the spin axis of possible exomoons of similar planets, I need estimates of their dimensionless quadrupole mass moment $J_2^\mathrm{p}$ and their spin angular momentum $S_\mathrm{p}$.

I will consider a gaseous giant planet with the same characteristics of, say, HD 106906b, whose relevant physical parameters are listed in Table\,\ref{tab6}. According to known formulas retrievable in, e.g.,\citep[][Chapter 4]{2000ssd..book.....M}, \citep[][Eq.\,(3)]{2011Icar..216..440H}, and \citep[][Appendix A]{2012Icar..217...77L}, its $J_2^\mathrm{p}$ and $S_\mathrm{p}$
\begin{table*}
\caption{Mass $M_\mathrm{p}$, equatorial radius $R_\mathrm{p}$, spinning period $P_\mathrm{p}$, dimensionless quadrupole mass moment $J_2^\mathrm{p}$, and spin angular momentum $S_\mathrm{p}$ of the gaseous giant planet HD109906b. The values of $M_\mathrm{p},\,R_\mathrm{p},\,P_\mathrm{p}$ were retrieved from Table\,3 of \citet{2020ApJ...905...37B}, while $J_2^\mathrm{p},\,S_\mathrm{p}$ were calculated following, e.g., \citet{2004A&A...428..691B,2009ApJ...698.1778R,2011A&A...528A..41L} by assuming $k_2^\mathrm{p}=0.52$ \citep{2009ApJ...698.1778R} for the planet's Love number. Here, $M_\mathrm{J},\,R_\mathrm{J},\,J_2^\mathrm{J},\,S_\mathrm{J}$ are referred to Jupiter. }
\label{tab6}
\centering
\begin{tabular}{ccccc}
\noalign{\smallskip}
\hline
$M_\mathrm{p}$ ($M_\mathrm{J}$) & $R_\mathrm{p}$ ($R_\mathrm{J}$) & $P_\mathrm{p}$ ($\mathrm{hr}$) & $J_2^\mathrm{p}$ ($J_2^\mathrm{J}$) & $S_\mathrm{p}$ ($S_\mathrm{J}$)\\
\hline
$11$ & $1.56$ & $4$ & $2.2$ & $43.2$ \\
\hline
\end{tabular}
\end{table*}
turn out to be
\begin{align}
0.5 \lb{jei2} &\lesssim \rp{J_2^\mathrm{p}}{J_2^\mathrm{J}}\lesssim 2.5,\\ \nonumber \\
27 \lb{jei} &\lesssim \rp{S_\mathrm{p}}{S_\mathrm{J}}\lesssim 45.
\end{align}
For Jupiter, it is $J_2^\mathrm{J}=0.0146966$ \citep{2018Natur.555..220I}, and $S_\mathrm{J} = 6.9\times 10^{38}\,\mathrm{J\,s}$ \citep{2003AJ....126.2687S}.
The ranges of values in \rfrs{jei2}{jei} were obtained for  $0.1\lesssim k_2^\mathrm{p}\lesssim 0.6$ \citep{2009ApJ...698.1778R}, where $k_2^\mathrm{p}$ is the planetary Love number  \citep{1939MNRAS..99..451S,1959cbs..book.....K,2009ApJ...698.1778R,2011A&A...528A..41L}.

\begin{figure}[H]
\centering
\centerline{
\vbox{
\begin{tabular}{c}
\epsfxsize= 16 cm\epsfbox{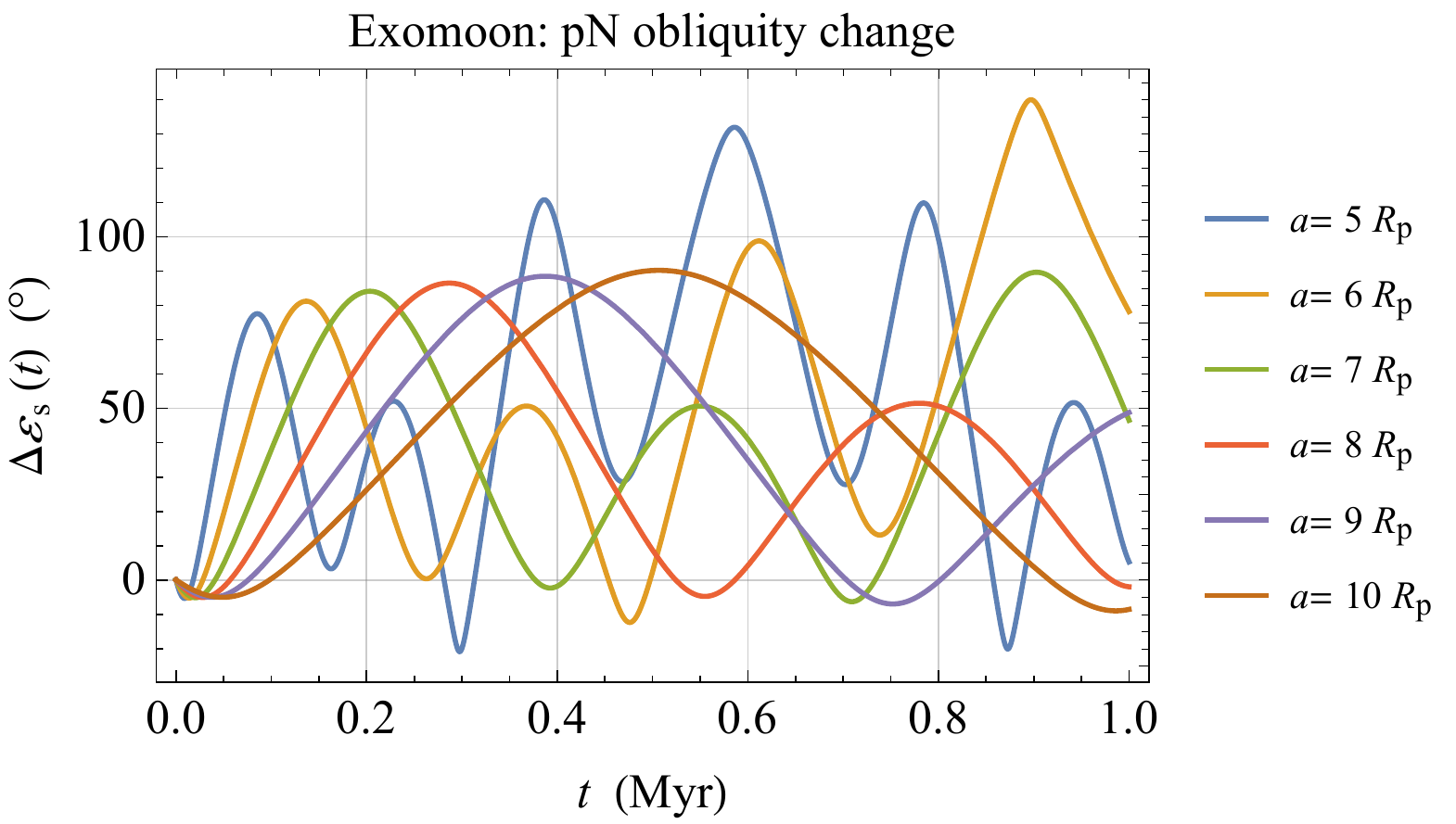}\\
\end{tabular}
}
}
\caption{
Numerically produced time series $\Delta\varepsilon_\mathrm{s}\ton{t} = \varepsilon_\mathrm{s}\ton{t}-\varepsilon_\mathrm{s}^0$, in $^\circ$, of the general relativistic pN variation of the obliquity $\varepsilon_\mathrm{s}$ to the ecliptic plane of a putative exomoon orbiting a gaseous giant planet with the same physical properties listed in Table\,\ref{tab6}. They were obtained by simultaneously integrating  the orbit-averaged \rfrs{one}{two} and \rfrs{IJ2}{OJ2}  for the rates of change of $\varepsilon_\mathrm{s},\,\alpha_\mathrm{s},\,\Omega,\,I$ over $1\,\mathrm{Myr}$.  The initial spin-orbit configuration, common to all the runs, is listed in Table\,\ref{tab5}.
}\label{fig6}
\end{figure}

By using the same initial conditions used in Section\,\ref{Jupi}, I numerical integrated \rfrs{one}{two} along with \rfrs{IJ2}{OJ2} for the planet of Table\,\ref{tab6}, subsequently confirmed by the integration of the spin and orbit precessional equations in vectorial form by \citet{1975PhRvD..12..329B}. As expected from the discussion in Section\,\ref{referee}, the resulting time series are essentially identical to those of
Figure\,\ref{fig1}\,to\,\ref{fig5}, apart from the characteristic timescales which are about one order of magnitude shorter amounting to $\simeq 0.07\,\mathrm{Myr}$. Indeed, the de Sitter frequency of HD109906b is just 11 times larger than that of Jupiter, being the planetocentric distances the same.
Just as an example, in Figure\,\ref{fig6}, I show the signatures obtained by varying $a$ from $5\,R_\mathrm{p}$ to $10\,R_\mathrm{p}$ for the same initial spin-orbit configuration of Table\,\ref{tab5}. Its resemblance with Figure\,\ref{fig5} is remarkable, with the exception of the timescale on the horizontal axis which, in this case, is about 10 times shorter.
\section{\textcolor{black}{The impact of the planetary quadrupole mass moment}}\lb{obls}
\textcolor{black}{
Here, I will look at the impact of the exomoon's own oblateness $J_2^\mathrm{s}$ on its spin rate.
}

\textcolor{black}{
According to, e.g., Eq.\,(47) of \citet{1975PhRvD..12..329B}, the Newtonian satellite's averaged spin rate due to the gravitational pull of the planet on its equatorial bulge is, in vectorial form,
\eqi
\dert{{\bds{\hat{S}}}_\mathrm{s}}{t} = {\mathbf{\Omega}}^\mathrm{s}_\mathrm{obl}\bds\times{\bds{\hat{S}}}_\mathrm{s},\lb{palla}
\eqf
with
\eqi
{\mathbf{\Omega}}_\mathrm{obl}^\mathrm{s} = \rp{\mu_\mathrm{p}\,M_\mathrm{s}\,J_2^\mathrm{s}\,R_\mathrm{s}^2}{2\,S_\mathrm{s}\,a^3\,\ton{1-e^2}^{3/2}}\,\qua{{\bds{\hat{S}}}_\mathrm{s}-3\ton{{\bds{\hat{S}}}_\mathrm{s}\bds\cdot\bds{\hat{L}}}\,\bds{\hat{L}}}.\lb{barba}
\eqf
From \rfrs{palla}{barba} along with \rfrs{Sx}{dazdt} and \rfrs{Jx}{Jz},  one obtains
\begin{align}
\dert{\varepsilon_\mathrm{s}}{t} \lb{plan1} & = -\rp{3\,\mu_\mathrm{p}\,J_2^\mathrm{s}\,M_\mathrm{s}\,R_\mathrm{s}^2\,\sin I\,\cos\zeta_\mathrm{s} }{2\,S_\mathrm{s}\,a^3\,\ton{1-e^2}^{3/2}}\,\qua{\cos\zeta_\mathrm{s} \,\sin I\,\ton{-\cos I\,\cos\varepsilon_\mathrm{s} + \sin I\,\sin\varepsilon_\mathrm{s}\,\sin\zeta_\mathrm{s} }},\\ \nonumber \\
\dert{\alpha_\mathrm{s}}{t} \lb{plan2} & = -\rp{3\,\mu_\mathrm{p}\,J_2^\mathrm{s}\,M_\mathrm{s}\,R_\mathrm{s}^2\,\csc\varepsilon_\mathrm{s}}{4\,S_\mathrm{s}\,a^3\,\ton{1-e^2}^{3/2}}\,\qua{\cos 2 \varepsilon_\mathrm{s}\,\sin 2 I\,\sin\zeta_\mathrm{s}  +\,\sin 2 \varepsilon_\mathrm{s}\,\ton{\cos^2 I - \sin^2 I\,\sin^2\zeta_\mathrm{s} }},
\end{align}
It turns out that, if ${\bds{\hat{S}}}_\mathrm{s}$ and $\bds{\hat{L}}$ are aligned, i.e. for $I=\varepsilon_\mathrm{s},\,\zeta_\mathrm{s} =-90^\circ$, \rfrs{plan1}{plan2} vanish. Now, also $I$ and $\Omega$ vary because of $J_2^\mathrm{s}$, as shown, e.g., by Eq.\,(64), Eq.\,(66), Eq.\,(71), and Eq.\,(72) of \citet{1975PhRvD..12..329B} or by  Eq.\,(7), Eq.\,(11), and Eq.\,(12) of \citet{2011CeMDA.111..105C}, and of $S_\mathrm{s}$ according to Eq.\,(64), Eq.\,(66), Eq.\,(68), and Eq.\,(69) of \citet{1975PhRvD..12..329B} in addition to the changes due to $J_2^\mathrm{s},\,S_\mathrm{p}$.
}

\textcolor{black}{
In order to include also \rfrs{plan1}{plan2} in the numerical integrations of Sections\,\ref{Jupi}\,to\,\ref{Big}, I need quantitative estimates of $J_2^\mathrm{s},\,S_\mathrm{s}$.
The quadrupole mass moment of the exomoon, assumed in hydrostatic equilibrium, can be calculated as \citep{2013ApJ...767..128C}
\eqi
J^\mathrm{s}_2=\rp{k^\mathrm{s}_2}{3}\ton{q^\mathrm{s}_c - \rp{q^\mathrm{s}_t}{2}},
\eqf
where
\eqi
q^\mathrm{s}_c\doteq \rp{\omega_\mathrm{s}^2\,R_\mathrm{s}^3}{\mu_\mathrm{s}}
\eqf
is due to the centrifugal acceleration felt by the satellite, whose gravitational parameter is $\mu_\mathrm{s}\doteq G\,M_\mathrm{s}$, spinning at the rate $\omega_\mathrm{s}$, while
\eqi
q^\mathrm{s}_t\doteq -3\,\ton{\rp{R_\mathrm{s}}{a}}^3\,\rp{M_\mathrm{p}}{M_\mathrm{s}}
\eqf
is due to tides raised by the nearby planet.
The parameter $k_2^\mathrm{s}$ is the exmoon's Love number which, for a homogeneous solid body, can be calculated as \citep{2000ssd..book.....M}
\eqi
k_2^\mathrm{s}\simeq \rp{3}{2\ton{1 + \rp{19\,\xi_\mathrm{s}}{2\,\rho_\mathrm{s}\,g_\mathrm{s}\,R_\mathrm{s}}}}.\lb{k2s}
\eqf
In \rfr{k2s}, $\rho_\mathrm{s}$ is the satellite's mean mass density,
\eqi
g_\mathrm{s}\simeq \rp{\mu_\mathrm{s}}{R_\mathrm{s}^2}
\eqf
is the exomoon's acceleration of gravity at its surface, and $\xi_\mathrm{s}$ is the rigidity of the satellite's solid core assumed homogeneous and incompressible; for a solid body, it is approximately \citep{2000ssd..book.....M}
\eqi
\xi_\mathrm{s}\simeq 5\times 10^{10}\,\mathrm{N\,m}^{-2}.
\eqf
The spin angular momentum $S_\mathrm{s}$ can be calculated as
\eqi
S_\mathrm{s} = j_\mathrm{s}\,M_\mathrm{s}\,R_\mathrm{s}^2\,\omega_\mathrm{s},
\eqf
where $j_\mathrm{s}$ is the normalized moment of inertia (NMoI) which, for a homogenous solid body amounts to $2/5=0.4$, being smaller for a differentiated internal structure with an inner compact core \citep{2015plsc.book.....D}.
In the following, I will adopt the values of the Earth's mass $M_\oplus$, radius $R_\oplus$, mean density $\rho_\oplus$, and NMoI $j_\oplus$  \citep{2010ITN....36....1P} for $M_\mathrm{s},\,R_\mathrm{s},\,\rho_\mathrm{s},\,j_\mathrm{s}$, while for the exomoon's angular speed I will assume
\eqi
\omega_\mathrm{s}=1.1\,\nk,
\eqf
corresponding to an approximate synchronization with the planetocentric orbital mean motion. Thus, it is
\begin{align}
0.1 \lb{Sei2} &\lesssim \rp{J_2^\mathrm{s}}{J_2^\oplus}\lesssim 0.9,\\ \nonumber \\
0.3 \lb{Sei} &\lesssim \rp{S_\mathrm{s}}{S_\oplus}\lesssim 0.8
\end{align}
for $5\,R_\mathrm{p}\leq a\leq 10\,R_\mathrm{p}$.
}

\textcolor{black}{
By defining
\begin{align}
\gamma^\mathrm{s}_\mathrm{obl} &\doteq \rp{3\,\mu_\mathrm{p}\,J_2^\mathrm{s}\,M_\mathrm{s}\,R_\mathrm{s}^2 }{2\,S_\mathrm{s}\,a^3\,\ton{1-e^2}^{3/2}}, \\ \nonumber \\
\Psi^\mathrm{dS}_\mathrm{obl}&\doteq \rp{\nu_\mathrm{dS}}{\gamma_\mathrm{obl}^\mathrm{s}} = \rp{S_\mathrm{s}\,\sqrt{a\,\ton{1-e^2}\,\mu_\mathrm{p}}}{c^2\,M_\mathrm{s}\,J_2^\mathrm{s}\,R_\mathrm{s}^2}, \\ \nonumber \\
\Pi^\mathrm{LT}_\mathrm{obl} &\doteq \rp{\nu_\mathrm{LT}}{\gamma_\mathrm{obl}^\mathrm{s}} =\rp{2\,S_\mathrm{p}\,S_\mathrm{s}}{3\,c^2\,M_\mathrm{p}\,M_\mathrm{s}\,J_2^\mathrm{s}\,R_\mathrm{s}^2},
\end{align}
it is possible to repeat the arguments of Section\,\ref{referee} applied, now, to the $J_2^\mathrm{s}$-driven rate of change as the fast one and to the pN precessions as slower components. Since for a circular orbit with $5\,R_\mathrm{p}\leq a\leq 10\,R_\mathrm{p}$ it is
\begin{align}
\Psi^\mathrm{dS}_\mathrm{obl}&\simeq 10^{-6}, \\ \nonumber \\
\Pi^\mathrm{LT}_\mathrm{obl} &\simeq 10^{-7}, 
\end{align}
it turns out that, over timescales much shorter than the characteristic pN ones, the exomoon's obliquity variations are driven by $J_2^\mathrm{s}$, and their amplitudes depend only on the planet-satellite initial spin-orbit configuration. The dependence on the planet's physical parameters and on the orbital radius is of the order of $\mathcal{O}\ton{\Psi^\mathrm{dS}_\mathrm{obl}},\,\mathcal{O}\ton{\Pi^\mathrm{LT}_\mathrm{obl}}$.
}

\begin{figure}[H]
\centering
\centerline{
\vbox{
\begin{tabular}{c}
\epsfxsize= 16 cm\epsfbox{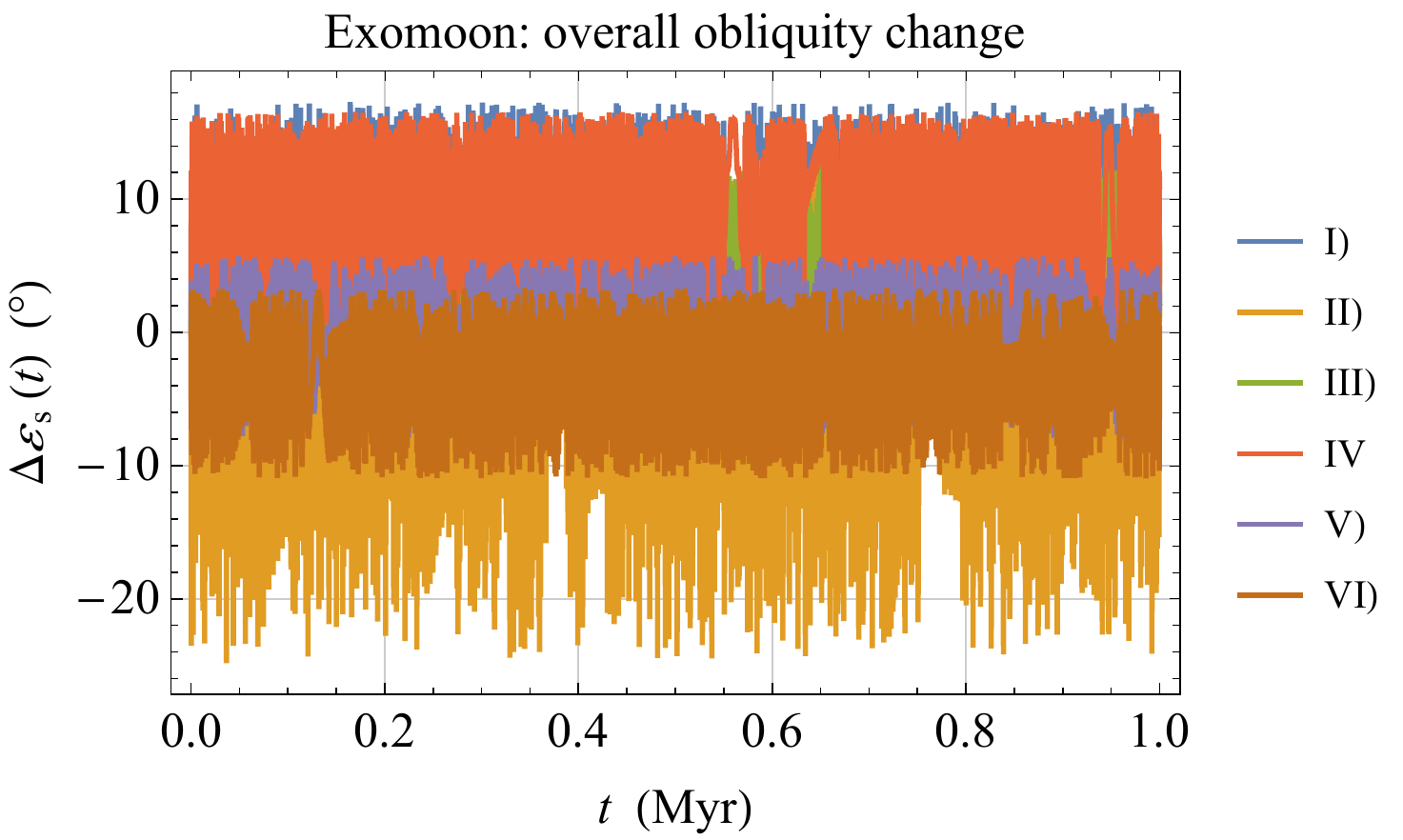}\\
\end{tabular}
}
}
\caption{
\textcolor{black}{
Numerically produced time series $\Delta\varepsilon_\mathrm{s}\ton{t} = \varepsilon_\mathrm{s}\ton{t}-\varepsilon_\mathrm{s}^0$, in $^\circ$, of the general relativistic pN variation of the obliquity $\varepsilon_\mathrm{s}$ to the ecliptic plane of a putative exomoon orbiting a gaseous giant planet with the same physical properties of Jupiter. They were obtained by simultaneously integrating  the orbit-averaged \rfrs{one}{two}, \rfrs{IJ2}{OJ2}, and \rfrs{plan1}{plan2}  for the rates of change of $\varepsilon_\mathrm{s},\,\alpha_\mathrm{s},\,\Omega,\,I$ over $1\,\mathrm{Myr}$.  The initial conditions, corresponding to a close alignment of $\bds{\hat{L}},\,{\bds{\hat{S}}}_\mathrm{p},\,{\bds{\hat{S}}}_\mathrm{s}$, are listed in Table\,\ref{tab1}. For the exomoon, the mass, radius, normalized moment of inertia and mean density of the Earth were adopted, while $\omega_\mathrm{s}=1.1\,\nk$ was assumed for its angular speed.
}
}\label{fig7}
\end{figure}

\textcolor{black}{
Figures\,\ref{fig7}\,to\,\ref{fig12}, corresponding to the same spin-orbit configurations and the same planet-satellite physical and orbital parameters of Figures\,\ref{fig1}\,to\,\ref{fig6}, were obtained by simultaneously integrating  \rfrs{one}{two}, \rfrs{plan1}{plan2}, and \rfrs{IJ2}{OJ2}. It turns out that the exomoon's oblateness $J_2^\mathrm{s}$ introduces high-frequency signatures which superimpose to the pN ones without canceling them.
Also in this case, I  successfully confirmed such results by numerically integrating the spin rate equations in vectorial form by \citet{1975PhRvD..12..329B}, whose resulting signatures agree with those in Figures\,\ref{fig7}\,to\,\ref{fig12} to within a sub-degree level.
}
\begin{figure}[H]
\centering
\centerline{
\vbox{
\begin{tabular}{c}
\epsfxsize= 16 cm\epsfbox{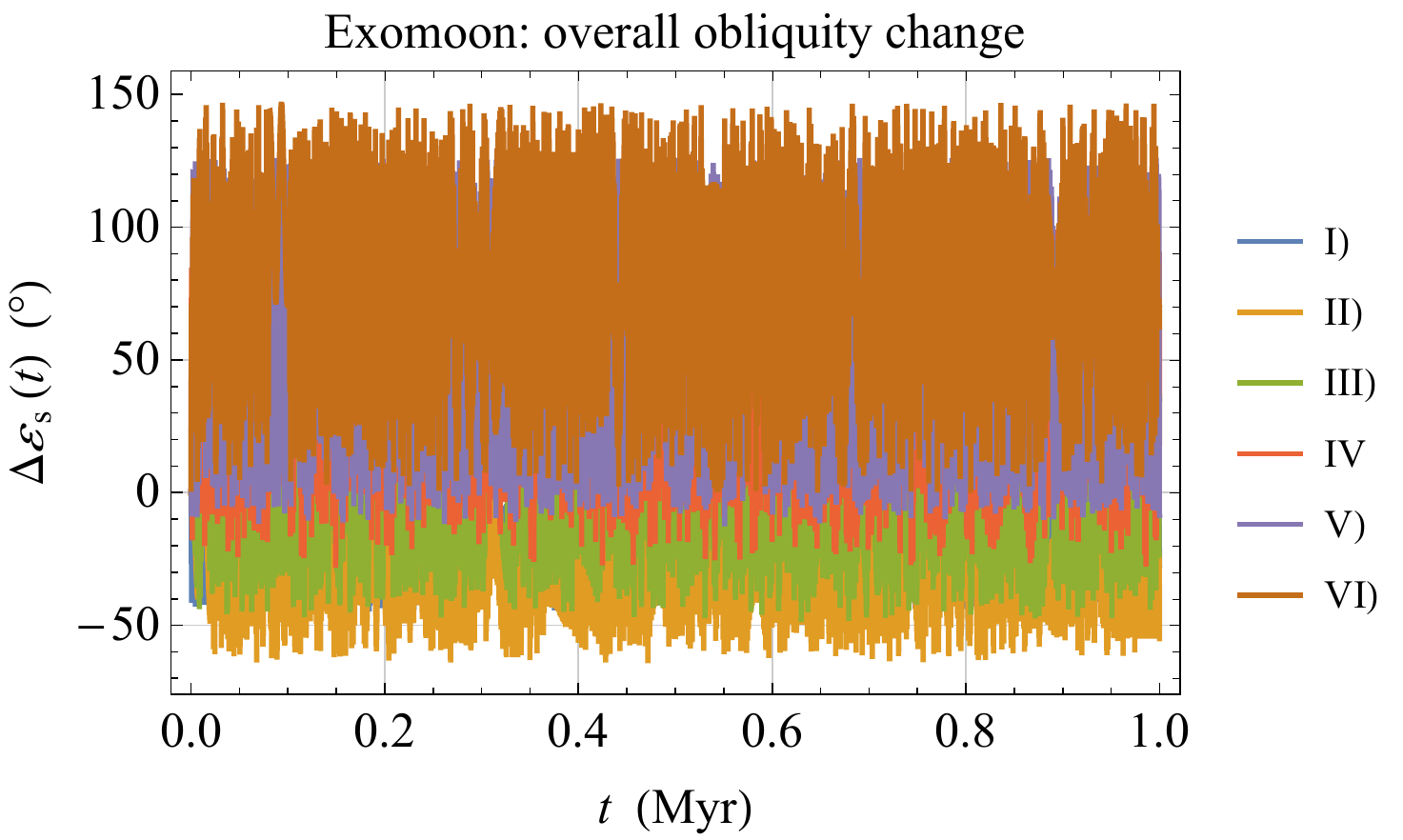}\\
\end{tabular}
}
}
\caption{
\textcolor{black}{
Numerically produced time series $\Delta\varepsilon_\mathrm{s}\ton{t} = \varepsilon_\mathrm{s}\ton{t}-\varepsilon_\mathrm{s}^0$, in $^\circ$, of the general relativistic pN variation of the obliquity $\varepsilon_\mathrm{s}$ to the ecliptic plane of a putative exomoon orbiting a gaseous giant planet with the same physical properties of Jupiter. They were obtained by simultaneously integrating  the orbit-averaged \rfrs{one}{two}, \rfrs{IJ2}{OJ2}, and \rfrs{plan1}{plan2}  for the rates of change of $\varepsilon_\mathrm{s},\,\alpha_\mathrm{s},\,\Omega,\,I$ over $1\,\mathrm{Myr}$.  The initial conditions, corresponding to $\bds{\hat{L}},\,{\bds{\hat{S}}}_\mathrm{p},\,{\bds{\hat{S}}}_\mathrm{s}$ lying almost in the same azimuthal plane but tilted differently from each other, are listed in Table\,\ref{tab2}. For the exomoon, the mass, radius, normalized moment of inertia and mean density of the Earth were adopted, while $\omega_\mathrm{s}=1.1\,\nk$ was assumed for its angular speed.
}
}\label{fig8}
\end{figure}
\begin{figure}[H]
\centering
\centerline{
\vbox{
\begin{tabular}{c}
\epsfxsize= 16 cm\epsfbox{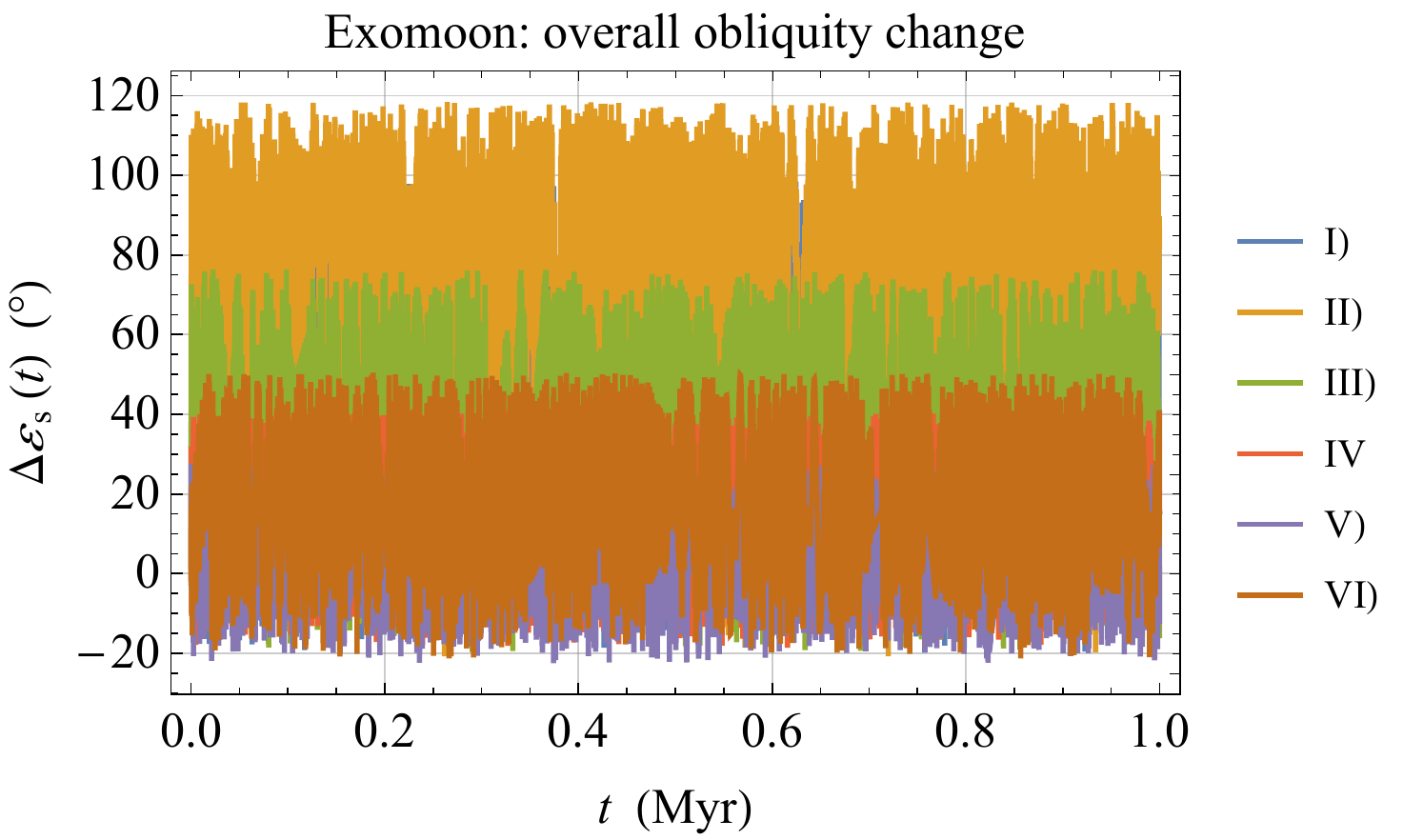}\\
\end{tabular}
}
}
\caption{
\textcolor{black}{
Numerically produced time series $\Delta\varepsilon_\mathrm{s}\ton{t} = \varepsilon_\mathrm{s}\ton{t}-\varepsilon_\mathrm{s}^0$, in $^\circ$, of the general relativistic pN variation of the obliquity $\varepsilon_\mathrm{s}$ to the ecliptic plane of a putative exomoon orbiting a gaseous giant planet with the same physical properties of Jupiter. They were obtained by simultaneously integrating  the orbit-averaged \rfrs{one}{two}, \rfrs{IJ2}{OJ2}, and \rfrs{plan1}{plan2}  for the rates of change of $\varepsilon_\mathrm{s},\,\alpha_\mathrm{s},\,\Omega,\,I$ over $1\,\mathrm{Myr}$.  The initial conditions, corresponding to $\bds{\hat{L}},\,{\bds{\hat{S}}}_\mathrm{p},\,{\bds{\hat{S}}}_\mathrm{s}$ tilted almost identically to the ecliptic but located in different azimuthal planes, are listed in Table\,\ref{tab3}. For the exomoon, the mass, radius, normalized moment of inertia and mean density of the Earth were adopted, while $\omega_\mathrm{s}=1.1\,\nk$ was assumed for its angular speed.
}
}\label{fig9}
\end{figure}
\begin{figure}[H]
\centering
\centerline{
\vbox{
\begin{tabular}{c}
\epsfxsize= 16 cm\epsfbox{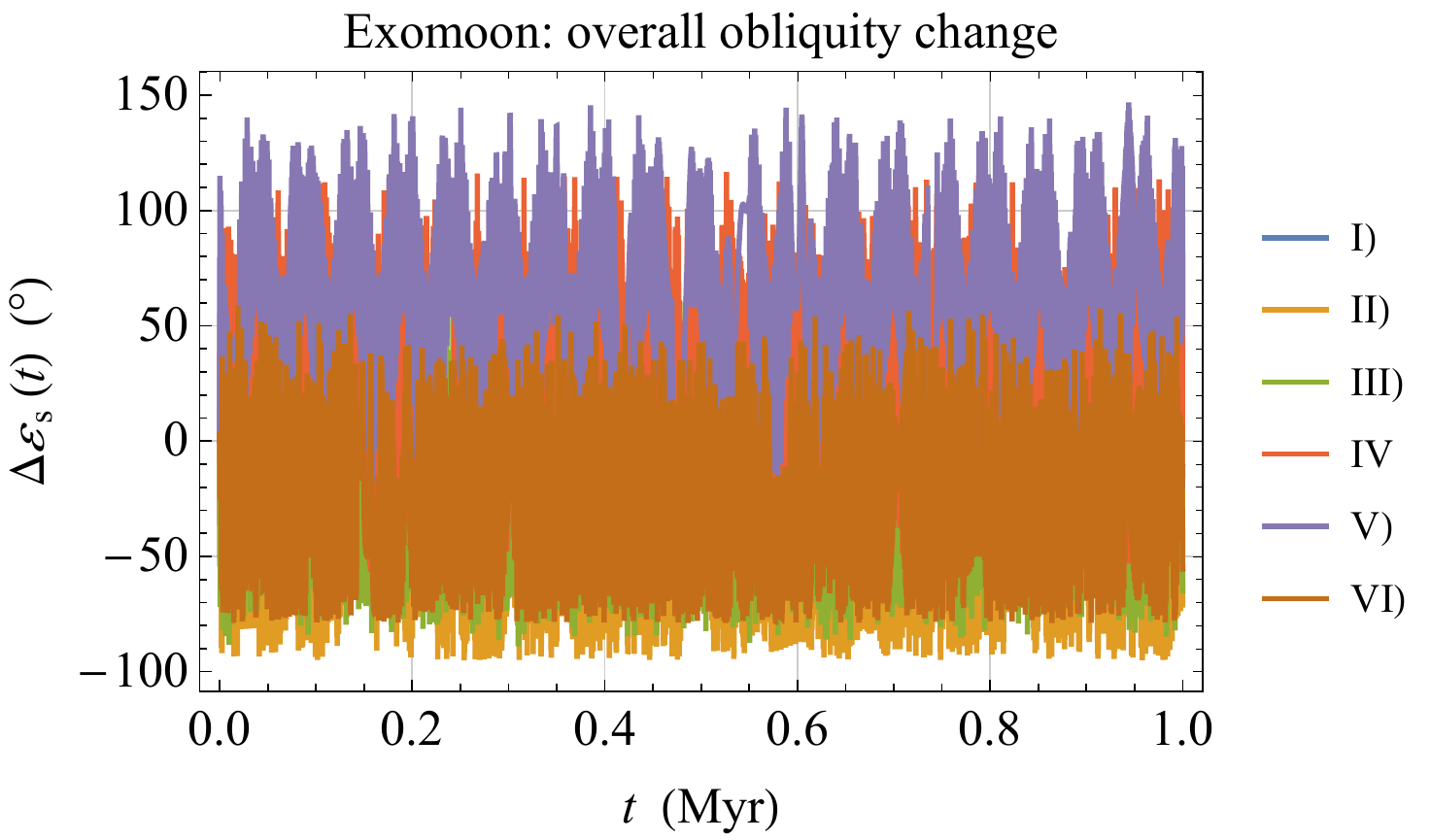}\\
\end{tabular}
}
}
\caption{
\textcolor{black}{
Numerically produced time series $\Delta\varepsilon_\mathrm{s}\ton{t} = \varepsilon_\mathrm{s}\ton{t}-\varepsilon_\mathrm{s}^0$, in $^\circ$, of the general relativistic pN variation of the obliquity $\varepsilon_\mathrm{s}$ to the ecliptic plane of a putative exomoon orbiting a gaseous giant planet with the same physical properties of Jupiter. They were obtained by simultaneously integrating  the orbit-averaged \rfrs{one}{two}, \rfrs{IJ2}{OJ2}, and \rfrs{plan1}{plan2}  for the rates of change of $\varepsilon_\mathrm{s},\,\alpha_\mathrm{s},\,\Omega,\,I$ over $1\,\mathrm{Myr}$.  The initial conditions, corresponding to $\bds{\hat{L}},\,{\bds{\hat{S}}}_\mathrm{p},\,{\bds{\hat{S}}}_\mathrm{s}$ arbitrarily oriented, are listed in Table\,\ref{tab4}. For the exomoon, the mass, radius, normalized moment of inertia and mean density of the Earth were adopted, while $\omega_\mathrm{s}=1.1\,\nk$ was assumed for its angular speed.}
}\label{fig10}
\end{figure}
\begin{figure}[H]
\centering
\centerline{
\vbox{
\begin{tabular}{c}
\epsfxsize= 16 cm\epsfbox{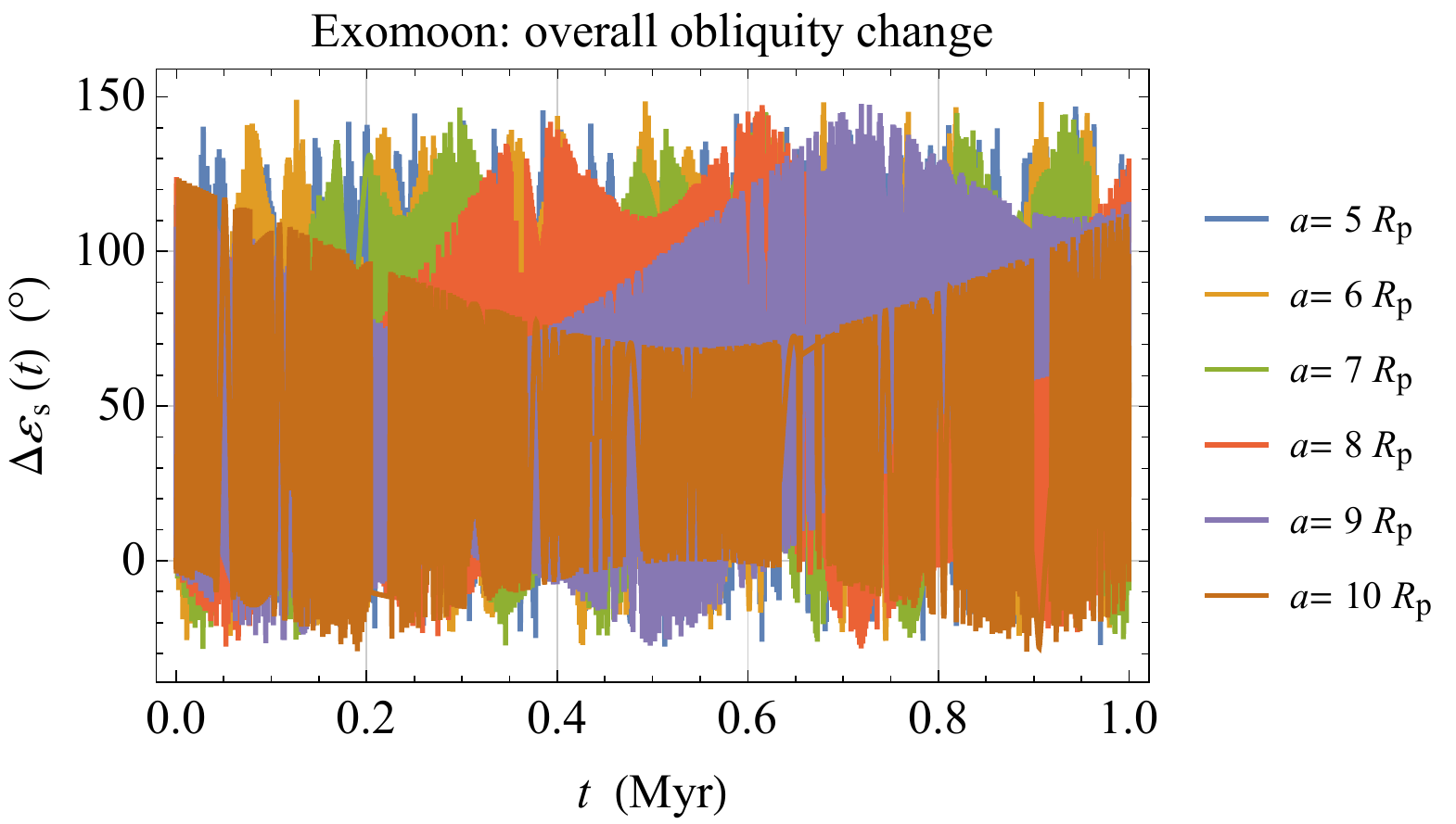}\\
\end{tabular}
}
}
\caption{
\textcolor{black}{
Numerically produced time series $\Delta\varepsilon_\mathrm{s}\ton{t} = \varepsilon_\mathrm{s}\ton{t}-\varepsilon_\mathrm{s}^0$, in $^\circ$, of the general relativistic pN variation of the obliquity $\varepsilon_\mathrm{s}$ to the ecliptic plane of a putative exomoon orbiting a gaseous giant planet with the same physical properties of Jupiter. They were obtained by simultaneously integrating  the orbit-averaged \rfrs{one}{two}, \rfrs{IJ2}{OJ2}, and \rfrs{plan1}{plan2}  for the rates of change of $\varepsilon_\mathrm{s},\,\alpha_\mathrm{s},\,\Omega,\,I$ over $50\,\mathrm{Myr}$.  The initial spin-orbit configuration, common to all the runs, is listed in Table\,\ref{tab5}. For the exomoon, the mass, radius, normalized moment of inertia and mean density of the Earth were adopted, while $\omega_\mathrm{s}=1.1\,\nk$ was assumed for its angular speed.
}
}\label{fig11}
\end{figure}
\begin{figure}[H]
\centering
\centerline{
\vbox{
\begin{tabular}{c}
\epsfxsize= 16 cm\epsfbox{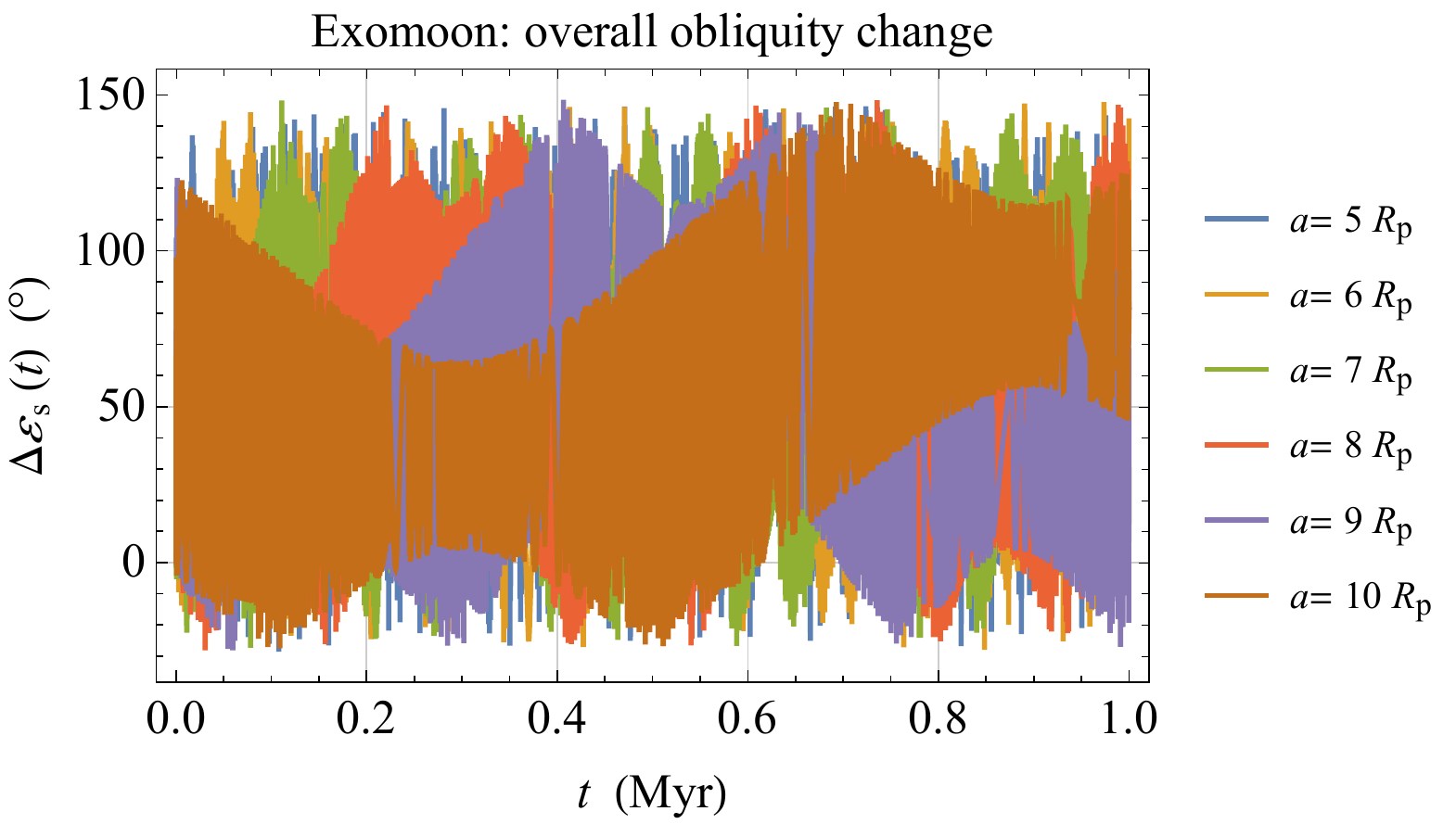}\\
\end{tabular}
}
}
\caption{
\textcolor{black}{
Numerically produced time series $\Delta\varepsilon_\mathrm{s}\ton{t} = \varepsilon_\mathrm{s}\ton{t}-\varepsilon_\mathrm{s}^0$, in $^\circ$, of the general relativistic pN variation of the obliquity $\varepsilon_\mathrm{s}$ to the ecliptic plane of a putative exomoon orbiting a gaseous giant planet with the same physical properties listed in Table\,\ref{tab6}. They were obtained by simultaneously integrating  the orbit-averaged \rfrs{one}{two}, \rfrs{IJ2}{OJ2}, and \rfrs{plan1}{plan2}  for the rates of change of $\varepsilon_\mathrm{s},\,\alpha_\mathrm{s},\,\Omega,\,I$ over $50\,\mathrm{Myr}$.  The initial spin-orbit configuration, common to all the runs, is listed in Table\,\ref{tab5}.  For the exomoon, the mass, radius, normalized moment of inertia and mean density of the Earth were adopted, while $\omega_\mathrm{s}=1.1\,\nk$ was assumed for its angular speed.
}
}\label{fig12}
\end{figure}
\section{\textcolor{black}{The impact of the distant star and the obliquity to the planetocentric orbital plane}}\lb{other}
\textcolor{black}{
Until now, I kept the ecliptic plane fixed. Actually, the distant star S, assumed here pointlike and with the same mass of our Sun, does affect the orbital angular momentum $\bds W$ of the revolution of the planet-satellite system $\mathcal{S}$ about it by means of three torques proportional to its mass $M_\mathrm{S}$ two of which depend on $J_2^\mathrm{s},\,J_2^\mathrm{p}$ \citep{2011CeMDA.111..105C}. Here, I investigate its possible influence on $\varepsilon_\mathrm{s}=\arccos\ton{ {\bds{\hat{S}}}_\mathrm{s}\bds\cdot\bds{\hat{W}} }$ by simultaneously integrating Eqs.\,(7)-(10) by \citep{2011CeMDA.111..105C} with the addition of the pN rates over the same time span of the figures in the previous Sections. I found that the inclusion of the torques by $M_\mathrm{S}$ does not substantially impact $\Delta\varepsilon_\mathrm{s}$.
}

\textcolor{black}{
As far as the obliquity $\vartheta_\mathrm{s}\doteq\arccos\ton{ {\bds{\hat{S}}}_\mathrm{s}\bds\cdot\bds{\hat{L}}}$ of the satellite's spin to the planetocentric orbital plane, which may play a role in the exomoon's habitability because of the energy output from the planet itself \citep{2013AsBio..13...18H}, its variations $\Delta\vartheta_\mathrm{s}$  are of no concern since their numerically integrated time series turn out to have negligible amplitudes.
}
\section{Summary and Conclusions}\lb{conclu}
I analytically and numerically studied the pN de Sitter and Lense-Thirring precessions of the spin of a spherically symmetric gyroscope freely moving in the deformed stationary spacetime of a massive rotating body for an arbitrary spin-orbit configuration. I applied my results to a putative exomoon orbiting different Jupiter-like gaseous giant planets, assumed to be at 1 au from a Sun-type main sequence star. In particular, I looked at the long-term pN variations $\Delta\varepsilon_\mathrm{s}$ of the satellite's obliquity $\varepsilon_\mathrm{s}$ \textcolor{black}{to the ecliptic plane, assumed fixed,} with respect to its initial value $\varepsilon_\mathrm{s}^0$. Indeed, the axial tilt is a key parameter in constraining the capability of hosting and sustaining life over long time spans since it controls the insolation received directly from the star at a given body's latitude. Thus, fast and large temporal changes of $\varepsilon_\mathrm{s}$, like those I found, may likely impact exomoons' habitability. Its detailed investigation from a climatological and planetological perspective is, however, outside the scopes of the present study.

First, I analytically derived orbit-averaged equations for the pN rates of change of the satellite's spin obliquity $\varepsilon_\mathrm{s}$ and azimuthal angle $\alpha_\mathrm{s}$ along with the equations for the precessions of the orbital inclination $I$ and longitude of ascending node $\Omega$  driven by both the classical quadrupole $J_2^\mathrm{p}$ and the  pN Lense-Thirring spin  component of the planetary gravitational field; the latter ones enter only indirectly $\mathrm{d}{\varepsilon_\mathrm{s}}/\mathrm{d}t$ and $\mathrm{d}{\alpha_\mathrm{s}}/\mathrm{d}t$.  By initially neglecting the satellite's oblateness $J_2^\mathrm{s}$ which induces its own direct spin precession, its spin obliquity rate is, thus, purely pN.
% and, for the considered scenario, mainly gravitoelectric, being the
%gravitomagnetic Lense-Thirring component 2-3 orders of magnitude smaller. Moreover, the orbital precessions of $I$ and $\Omega$ are dominated by the %planet's oblateness by 7 orders of magnitude with respect to the Lense-Thirring ones. Even with such approximations, an exact analytical solution of the %resulting system of equations for $\varepsilon_\mathrm{s},\,\alpha_\mathrm{s},\,I,\,\Omega$ cannot  be found for an arbitrary spin-orbit configuration. %Nonetheless, some qualitative features of $\varepsilon_\mathrm{s}\ton{t}$ can still be inferred. In general, it represents a non-harmonic signal with %complex temporal pattern whose characteristic timescale is primarily determined by the de Sitter frequency. Furthermore, its amplitude is independent of %the planet's physical properties and the exomoon's planetocentric distance, depending only on the initial spin-orbit configuration.

Subsequently, the equations for $\varepsilon_\mathrm{s},\,\alpha_\mathrm{s},\,I,\,\Omega$ were numerically integrated over $1\,\mathrm{Myr}$.  I started by varying the spin-orbit configuration by keeping the planetocentric distance from a Jupiter-like planet fixed to $5\,R_\mathrm{p}$.  I considered four different scenarios. First, I assumed $\bds{\hat{L}},\,{\bds{\hat{S}}}_\mathrm{p},\,{\bds{\hat{S}}}_\mathrm{s}$ almost aligned with each other up to a few degrees: the ideal condition of perfect alignment of the three angular momenta would imply the absence of any spin precessions. I found the resulting ranges of variation $\varepsilon_\mathrm{s}^\mathrm{max}-\varepsilon_\mathrm{s}^\mathrm{min}$ large enough to be likely significant for habitability, amounting to $\simeq 3^\circ-17^\circ$. Then, I considered the cases in which $\bds{\hat{L}},\,{\bds{\hat{S}}}_\mathrm{p},\,{\bds{\hat{S}}}_\mathrm{s}$ a) Share almost the same azimuthal plane but are differently tilted to the ecliptic b) Are located in different azimuthal planes with almost the same axial tilts c) Are arbitrarily oriented in space. In all such cases, $\Delta\varepsilon_\mathrm{s}$ experiences relevant variations up to tens and, sometimes, even hundreds degrees. All such integrations, covering $1\,\mathrm{Myr}$, exhibit non-harmonic temporal patterns and a characteristic timescale of about $0.7\,\mathrm{Myr}$.
Finally, I made numerical integrations by varying \textcolor{black}{also} the exmoon's planetocentric distance from $5\,R_\mathrm{p}$ to $10\,R_\mathrm{p}$ for a given spin-orbit configuration. I found that all the curves for $\Delta\varepsilon_\mathrm{s}$ retain essentially the same amplitudes, showing increasing characteristic timescales with distance.

I successfully tested my results with another set of runs over the same time spans  based on the vectorial form of the spin and orbital precessions retrieved in the literature. The resulting times series agree with the previous ones up to less than a degree over the time span adopted.

I looked also at another parent planet for which the physical parameters of the existing exoplanet HD109906b were used. In particular, its mass is 11 times larger than that of Jupiter, and its oblateness $J_2^\mathrm{p}$ and spin angular momentum $S_\mathrm{p}$ can be up to $2.5$ and $45$ times larger than the Jovian ones. The temporal patterns and the magnitudes of the resulting time series, calculated with the same sets of initial conditions, are similar to those for a Jupiter-like body, but their characteristic timescale is about the order of magnitude shorter. Indeed, the de Sitter frequency of a HD109906b-type planet is just about 10 times larger than that of Jupiter for the same planetocentric distances.

Then, I studied the impact of the own oblateness $J_2^\mathrm{s}$ of the exomoon on $\varepsilon_\mathrm{s}$ by adding its Newtonian torque to the numerically integrated equations for the spin rates after having analytically worked out its spin rates as well. In calculating the satellite's quadrupole mass moment and spin angular momentum $S_\mathrm{s}$, I assumed the relevant physical parameters of the Earth by allowing for the exomoon's angular rotational speed $\omega_\mathrm{s}$ a value, say, $10\%$ larger than that of the planetocentric mean motion $\nk$. As a result, I obtained very high frequency signatures of comparable amplitudes to the pN ones to which they are superimposed without canceling them. Also in this case, the numerical integrations of the analytically worked out averaged spin rate equations agree with those performed with their vectorial form.

I tested the assumption that the plane of the astrocentric motion of the planet-satellite binary can be considered fixed by adding the torques arising from the action of the distant star to the dynamical model to be integrated. The resulting time series for $\Delta\varepsilon_\mathrm{s}$ do not noticeably differ from those previously obtained by keeping the ecliptic fixed.

Finally, I  investigated also the classical and pN variations of the obliquity of the exomoon's spin with respect to the plane of its orbit around the host planet, which may be another relevant factor in the total energy balance because of the irradiation from the planet itself due to both the reflected sunlight and the infrared radiation. It turned out to be negligible.

As directions for future work, a further effect which is likely worth of further investigations is the impact of the tidal torques which, among other things, tend to align the orbital angular momentum of the planetocentric orbit with the planet and satellite's spins in order to see if, and to which extent, they are effectively counterbalanced by the relativistic signatures obtained here, especially when their characteristic frequencies increase because of a heavier primary than Jupiter.

In conclusion, the pN temporal variations of the tilt of the exomoon's spin axis to the plane of the planet-satellite's orbit around a distant Sun-like main sequence star are fast and large enough to have most likely a significant impact on its habitability for a variety of different spin-orbit configurations, even when the high frequency modulation due to the satellite's own oblateness is taken into account.  It is true also in the  scenario in which the satellite's spin is almost aligned with the planetary one and with the planetocentric orbital angular momentum up to offsets of a few degrees.
%To my knowledge, it is the first time that it is shown that general relativity may directly have a macroscopic impact on life in a likely common %astronomical scenario.
Future climatological and planetological studies on the habitability of exomoons should include also such effects in the overall budget of the dynamical constraints to life sustainability.
\section*{Data availability}
No new data were generated or analysed in support of this research.
\bibliography{exopbib,PXbib}{}

\end{document}